\documentclass[conference,compsoc]{IEEEtran}
\ifCLASSOPTIONcompsoc
  \usepackage[nocompress]{cite}
\else
  \usepackage{cite}
\fi
\ifCLASSINFOpdf
\else
\fi
\usepackage{cite}
\usepackage{amsmath,amssymb,amsfonts}
\usepackage{algorithm}
\usepackage{algpseudocode}
\usepackage{graphicx}
\usepackage{enumitem}
\usepackage{bm}

\usepackage{textcomp}
\usepackage{xcolor}
\usepackage{stfloats}
\usepackage{enumitem}
\usepackage{multirow}
\usepackage{tikz,calc}
\usepackage{colortbl}
\usepackage{booktabs}
\usepackage{arydshln}
\usepackage[most]{tcolorbox}
\usepackage{pifont}
\usepackage{ulem}
\usepackage{subcaption}
\usepackage{xspace}
\usepackage{placeins}
\usepackage{hyperref}
\usepackage{xurl}

\newcommand{\NAME}{\texttt{FDLLM}\xspace}
\begin{document}
\tcbset{
  colback=gray!25!white,
  colframe=gray!25!white,
  breakable,
  boxrule=0mm,
  boxsep=0mm,
  left=2mm,
  right=2mm,
  top=2mm,
  bottom=2mm
}
%
\title{FDLLM: A Dedicated Detector for Black-Box LLMs Fingerprinting}

\author{
    \IEEEauthorblockN{
        Zhiyuan Fu\IEEEauthorrefmark{1}\IEEEauthorrefmark{2}\IEEEauthorrefmark{3}, 
        Junfan Chen\IEEEauthorrefmark{1}\IEEEauthorrefmark{2}\IEEEauthorrefmark{3}, 
        Lan Zhang\IEEEauthorrefmark{5}, 
        Ting Yang\IEEEauthorrefmark{4}\IEEEauthorrefmark{2}\IEEEauthorrefmark{3},
        Jun Niu\IEEEauthorrefmark{4}\IEEEauthorrefmark{2},
        Hongyu Sun\IEEEauthorrefmark{1}, \\
        Ruidong Li\IEEEauthorrefmark{3}, 
        Peng Liu\IEEEauthorrefmark{6}, 
        Jice Wang\IEEEauthorrefmark{1},
        Fannv He\IEEEauthorrefmark{1},
        Qiuling Yue\IEEEauthorrefmark{1},
        Yuqing Zhang\IEEEauthorrefmark{1}\IEEEauthorrefmark{2}\IEEEauthorrefmark{4}
    }

    \IEEEauthorblockA{\IEEEauthorrefmark{1}College of Cyberspace Security, Hainan University, Haikou, China}
    \IEEEauthorblockA{\IEEEauthorrefmark{2}National Computer Network Intrusion Protection Center, University of Chinese Academy of Sciences, Beijing, China}
    \IEEEauthorblockA{\IEEEauthorrefmark{3}Graduate School of Natural Science and Technology, Kanazawa University, Kanazawa, Japan}
    \IEEEauthorblockA{\IEEEauthorrefmark{4}School of Cyber Engineering, Xidian University, Xi'an, China}
    \IEEEauthorblockA{\IEEEauthorrefmark{5}School of Informatics, Computing, and Cyber Systems, Northern Arizona University, Flagstaff, Arizona, USA}
    \IEEEauthorblockA{\IEEEauthorrefmark{6}College of IST, Pennsylvania State University, USA}

    \IEEEauthorblockA{
        \{fuzy, chenjunfan\}@hainanu.edu.cn, moirai.zhang@gmail.com, yangt@nipc.org.cn,\\ niujun@stu.xidian.edu.cn, sunhy@hainanu.edu.cn,
        liruidong@ieee.org, pxl20@psu.edu,\\ \{wangjice, hefannv, yueqiuling\}@hainanu.edu.cn, zhangyq@ucas.ac.cn
    }
}


\maketitle

\begin{abstract}
Large Language Models (LLMs) are rapidly transforming the landscape of digital content creation.
However, the prevalent black-box Application Programming Interface (API) access to many LLMs introduces significant challenges in accountability, governance, and security. LLM fingerprinting, which aims to identify the source model by analyzing statistical and stylistic features of generated text, offers a potential solution. 
Current progress in this area is hindered by a lack of dedicated datasets and the need for efficient, practical methods that are robust against adversarial manipulations.
To address these challenges, we introduce \textit{FD-Dataset}, a comprehensive bilingual fingerprinting benchmark comprising 90,000 text samples from 20 famous proprietary and open-source LLMs. 
Furthermore, we present \NAME, a novel fingerprinting method that leverages parameter-efficient Low-Rank Adaptation (LoRA) to fine-tune a foundation model. 
This approach enables LoRA to extract deep, persistent features that characterize each source LLM. Through our analysis, we find that LoRA adaptation promotes the aggregation of outputs from the same LLM in representation space while enhancing the separation between different LLMs. This mechanism explains why LoRA proves particularly effective for LLM fingerprinting.
Extensive empirical evaluations on \textit{FD-Dataset} demonstrate \NAME's superiority, achieving a Macro F1 score 22.1\% higher than the strongest baseline. 
\NAME also exhibits strong generalization to newly released models, achieving an average accuracy of 95\% on unseen models. 
Notably, \NAME remains consistently robust under various adversarial attacks, including polishing, translation, and synonym substitution. Experimental results show that FDLLM reduces the average attack success rate from 49.2\% (LM-D) to 23.9\%.
\end{abstract}


%
\IEEEpeerreviewmaketitle

\section{Introduction}

The rapid proliferation of Large Language Models (LLMs), such as ChatGPT~\cite{openai2024gpt4technicalreport}, Claude~\cite{claude35haiku}, and Gemini~\cite{team2024gemini}, is fundamentally reshaping digital content creation. However, most LLMs are accessible only through proprietary and opaque Application Programming Interface (API), which introduces substantial challenges in terms of security, accountability, and governance~\cite{hou2024security, shen2025gptracker}. 

\textbf{Motivation.} In this context, the ability to identify the specific source model responsible for generating a given piece of text, commonly referred to as LLM fingerprinting, has become increasingly important. This capability supports key applications such as tracing the provenance of information, mitigating the spread of misinformation~\cite{aioutputdisclosure}, enforcing copyright protections~\cite{aicopyright}, and ensuring legal and ethical responsibility~\cite{mozes2023use, gosztonyi2025theory}.

A useful comparison can be drawn with traditional cybersecurity practices. During a security assessment, the initial phase often involves reconnaissance, where techniques such as operating systems (OS) fingerprinting~\cite{pasquini2024llmmap} are employed to infer the identity of a system based on observable behaviors and responses. In a similar manner, LLM fingerprinting seeks to identify the underlying model and version accessible through a black-box API by examining the statistical and stylistic characteristics present in its generated outputs. 
A central challenge in this evolving landscape is the lack of reliable attribution for LLM-generated text (LLMGT). 

Unlike OS, LLMs leave their fingerprints not in network signals but in the texts they generate. 
These fingerprints often take the form of implicit statistical or stylistic patterns, which are unintentionally embedded by each model during generation~\cite{abdali2024decoding}. 
Detecting and analyzing these patterns makes it possible to attribute a text to its source LLM, moving beyond simple AI-vs-human detection~\cite{yang2023dna,xu2024freqmark,Shi_2024} toward fine-grained identification of the specific model. However, existing methods for leveraging such fingerprints still face some limitations in practice.

Methods relying on easily discernible statistical features~\cite{gehrmann2019gltr,solaiman2019releasestrategiessocialimpacts,su2023detectllmleveraginglogrank}, such as those employed by sentiment classifiers that utilize high-frequency cues like sentiment words or entity markers, often prove susceptible to adversarial attacks.
In contrast, approaches based on carefully constructed prompt injection schemes~\cite{pasquini2024llmmap} tend to require frequent API queries, which limits their practicality.
Meanwhile, traditional model-based detectors~\cite{mitchell2023detectgpt,guo2023closechatgpthumanexperts,shi2024ten}, often built on less sophisticated architectures, are ill-equipped to handle the complexity and nuance of texts generated by advanced, contemporary LLMs, rendering them largely ineffective for robust attribution. These approaches are typically evaluated on a narrow set of models, languages, or domains and struggle to strike a balance between efficiency and generalization ability.

Addressing these issues requires overcoming three core challenges:

\noindent \textbf{Challenge 1 (C1):}
There is a lack of dedicated datasets for LLM fingerprinting. This gap limits progress in evaluating and improving attribution across different models, languages, and domains.

\noindent \textbf{Challenge 2 (C2):}
LLM fingerprinting methods need to be efficient and practical. They should be able to work efficiently without relying heavily on external APIs or ample computing resources and remain adaptable to new models.

\noindent \textbf{Challenge 3 (C3):}
Robust attribution remains challenging in the face of simple adversarial manipulations, such as translation or synonym substitution, which can obscure or alter the features used for reliable detection.

Collectively, these challenges motivate the key insights that underlie our proposed approach (see Section~\ref{sec:key_insights}).

\textbf{LLMs Fingerprinting Dataset.} 
To tackle \textbf{(C1)}, which highlights the lack of dedicated datasets for LLM fingerprinting, we introduce \textit{FD-Dataset}, a bilingual fingerprinting benchmark consisting of 90,000 text samples from 20 widely used LLMs, including both proprietary and open-source families.
By standardizing prompts and sampling conditions across all models, our triplet-based data collection helps ensure that observed differences are primarily attributable to model-specific generation patterns.
This design enables the reliable capture of both implicit and robust LLM fingerprints.

\textbf{LLM Fingerprinting Detection Framework.}
To address \textbf{(C2)}, which calls for practical and efficient fingerprinting methods, we present the \NAME (\textbf{F}ingerprint \textbf{D}etection for \textbf{L}arge \textbf{L}anguage \textbf{M}odels).
\NAME utilizes parameter-efficient Low-Rank Adaptation (LoRA) to fine-tune a foundation model, learning deep, persistent features that distinguish different source LLMs. Attribution is performed in a single forward pass without querying candidate LLM APIs.

\textbf{Attribution Mechanisms Analysis.}
An innovation of our approach is the repurposing of LoRA-based adaptation for attribution tasks. 
While conventional LoRA fine-tuning typically enhances performance on downstream tasks such as sentiment or topic classification~\cite{SWQBZZ25}, our method is designed to capture deeper, persistent features. These include rare word choices, tokenizer boundary behaviors, implicit punctuation patterns, and sampling noise. Such features often uniquely reflect the identity of the source LLM.
Visualization results suggest that LoRA adaptation encourages the clustering of outputs from the same model and more precise separation between different models, shedding light on its effectiveness for attribution.
Unlike explicit semantic cues used in standard Natural Language Processing tasks, these implicit patterns must remain stable even after heavy post-processing (e.g., translation or polishing), which makes robust fingerprinting particularly challenging. 
These unique challenges motivate the design specifics of \NAME.

\textbf{High Detection Performance.}
Through extensive empirical experiments on the \textit{FD-Dataset} we demonstrate that \NAME achieves a Macro F1 score 22.1\% higher than the strongest baseline (LM-D~\cite{ippolito2020automaticdetectiongeneratedtext}) on challenging tasks.
For newly released models, \NAME maintains high accuracy through incremental adaptation with only a small number of labeled samples. For example, the average accuracy on unseen models such as GPT-4.1 and Phi4 reaches 95\%.
Under out-of-distribution (OOD) scenarios, \NAME also performs strongly. On a challenging QA dataset containing both English and LLM-translated Chinese answers with high answer similarity across models, it achieves a Macro F1 score of \textbf{49.9\%}, outperforming all competing methods.

\textbf{Robustness of \NAME.}
To further address \textbf{(C3)}, we evaluate \NAME under three practical adversarial attack settings: translation, polishing, and synonym substitution. 
\NAME demonstrates consistent robustness: synonym substitution attacks result in an attack success rate below 7\%, polishing causes only a 27.3\% drop in F1, and even the challenging translation attack leads to an F1 degradation of 54.5\%, still outperforming all other methods.

\textbf{Contributions.} The contributions of this paper are as follows:

\begin{itemize}[leftmargin=1em]
\item We propose \NAME, a task-specific framework for fingerprinting LLMGT via LoRA-based fine-tuning. Extensive empirical results demonstrate that LoRA adaptation significantly improves attribution performance, achieving a Macro F1 score 22.1\% higher than the strongest baseline and maintaining robustness under realistic adversarial attacks.
\item We introduce \textit{FD-Dataset}, a large-scale, bilingual, and multi-domain benchmark comprising 90,000 samples from 20 widely used LLMs. 
This dataset is specifically designed to facilitate research in LLM attribution.
\item We empirically demonstrate that LoRA-based adaptation improves model attribution by increasing inter-class separation and reducing intra-class variance in the feature space.
\end{itemize}

\section{Background and Motivation}
\label{sec:background}

\subsection{Background}

\textbf{Decoder-only Architectures in LLMs.}
The majority of widely used LLMs, such as ChatGPT~\cite{openai2024gpt4technicalreport}, Claude~\cite{claude35haiku}, and Gemini~\cite{team2024gemini}, are built upon decoder-only architectures. This design is preferred because its pre-training strategy, autoregressive language modeling, is naturally suited for open-ended text generation tasks.  In light of the effectiveness and popularity of decoder-only architectures~\cite{peftguard}, we also select this structure as the backbone for our method.

\textbf{LoRA~\cite{hu2021lora}.}
A widely adopted parameter-efficient fine-tuning (PEFT) method that enables LLMs to adapt to downstream tasks with minimal additional overhead.
Modern LLMs typically consist of billions of parameters, making full fine-tuning computationally expensive and memory-intensive. LoRA addresses this challenge by injecting trainable low-rank matrices into the model’s architecture while keeping the original weights frozen. The core idea is that the parameter update space for many tasks lies in a much lower-dimensional subspace than the full parameter space of the model.

Formally, consider a linear transformation in the pre-trained model represented by a weight matrix $\mathbf{W} \in \mathbb{R}^{d \times k}$. Instead of updating $\mathbf{W}$ directly, LoRA re-parameterizes it as:
\begin{equation}
\text{LoRA}(\mathbf{W}) = \mathbf{W} + \Delta \mathbf{W},
\label{eq:lora}
\end{equation}
where the update matrix $\Delta \mathbf{W}$ is defined as:
\begin{equation}
\Delta \mathbf{W} = \frac{\alpha}{r} \mathbf{A} \mathbf{B},
\label{eq:loraw}
\end{equation}
with $\mathbf{A} \in \mathbb{R}^{d \times r}$ and $\mathbf{B} \in \mathbb{R}^{r \times k}$ denoting the trainable low-rank matrices, $r$ being the rank of the adaptation, and $\alpha$ a scaling factor.

The hyperparameters $r$ and $\alpha$ play a critical role in determining the expressive capacity and stability of the adaptation. A larger $r$ increases the number of trainable parameters, enhancing the model's ability to capture complex patterns and nuanced textual features. The scaling factor $\alpha$ controls the magnitude of the low-rank update, balancing the influence of $\Delta \mathbf{W}$ against the frozen base/backbone weights.

\begin{figure*}[ht]
    \centering
    \includegraphics[width=0.89\linewidth]{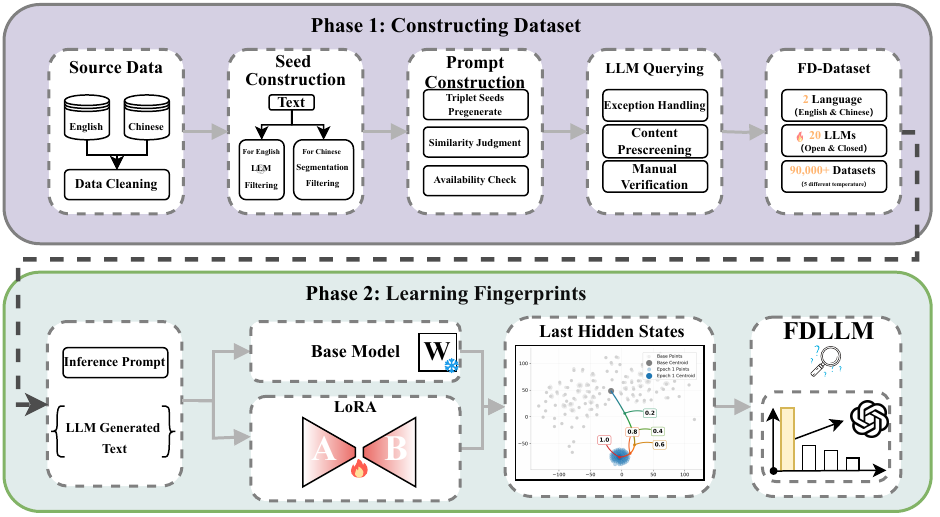}
    \caption{The overall framework of the article. The \NAME framework consists of two phases. \textbf{Phase 1: Constructing Dataset.} Seed prompts are built and cleaned in both English and Chinese, filtered using LLMs, and checked for availability to produce a large-scale multilingual dataset from 20 LLMs. \textbf{Phase 2: Learning Fingerprints.} Input text is evaluated by the LoRA fine-tuned \NAME model to extract discriminative features for fingerprinting detection.}
    \label{fig:framework}
\end{figure*}

\subsection{Key Insights Motivating Our Approach}
\label{sec:key_insights}
Our development of \NAME is guided by several key insights that address the fundamental limitations of current LLM attribution methods:

\noindent\textbf{Insight 1: Learning of distinctive fingerprints benefits from
 systematic multilingual and multi-domain data collection.}  
Some empirical analysis~\cite{abassy-etal-2024-llm,pudasaini2025benchmarking} reveals that LLM fingerprints vary significantly across languages and domains, owing to differences in model tokenization behavior, training corpus composition, and optimization objectives. This inconsistency poses a major challenge for attribution systems aiming for real-world deployment.

In particular, traditional approaches often rely on statistical methods or feature engineering. Some methods use language-specific tools, such as Snowball~\cite{Shi_2024}, in an attempt to generalize across languages. However, these approaches face limitations due to insufficient support for non-English tokenization or morphological analysis.

Moreover, for low-resource languages, the absence of robust data generation pipelines exacerbates the difficulty of building reliable attribution datasets. As a result, attribution models trained on homogeneous or monolingual datasets tend to under-represent the diverse fingerprinting signals, making it difficult to detect multilingual, multi-domain texts~\cite{macko2024authorship}.

\noindent\textbf{Insight 2: Foundation models, when adapted in a parameter-efficient manner, can learn robust fingerprints that remain stable under various transformations.}
Recent studies suggest that pre-trained foundation models have been shown to encode
implicit generative patterns that can distinguish outputs from different LLMs~\cite{hans2024spotting}. 

For instance, even in few-shot settings, models like Qwen2~\cite{iourovitski2024hide} or GPT-3.5~\cite{lu2023large} have shown competitive performance in attribution tasks. Conversely, full model fine-tuning~\cite{li2025responsible}, while effective in principle, is often computationally prohibitive and prone to catastrophic forgetting.

\noindent\textbf{Insight 3: Adversarial robustness requires learning deeper layer features and fingerprints that are resistant to semantic-preserving adversarial manipulations.}
Our investigation into existing attribution methods reveals a critical vulnerability: many existing approaches still rely heavily on statistical patterns~\cite{xu2024freqmark} or stylistic markers that are easily disrupted by simple textual modifications~\cite{he2024mgtbench, SWQBZZ25}. For example, traditional detectors based on shallow statistical features suffer substantial performance degradation when facing adversarial manipulations such as paraphrasing, synonym substitution, or random spacing. 

Through systematic evaluation, we found that even minor edits, such as word substitutions or sentence reordering~\cite{pasquini2024llmmap}, can significantly reduce the effectiveness of current fingerprinting techniques. This finding highlights the need for detection models that are more semantically grounded and aware of deeper representations.

\section{Design}

In this section, we describe the design of \NAME by introducing its framework and then outlining the two phases.
\subsection{Overall Framework}
The overall framework of \NAME is illustrated in Figure~\ref{fig:framework}, which outlines a two-phase general workflow.
\textit{Insight 1} highlights the importance of diverse, controlled data in identifying generation-specific behaviors. Section~\ref{sec:data_construction} presents a high-coverage dataset tailored to elicit model-unique fingerprints under varied prompts and temperatures (\textbf{Phase 1}).
\textit{Insight 2} emphasizes that attribution must encompass deeper, model-specific features. Section~\ref{sec:representation_adaptation} proposes \NAME, which applies LoRA-based, PEFT to a frozen backbone. Unlike conventional approaches that use LoRA for downstream tasks such as sentiment or topic classification, our method is specifically designed to capture deeper, persistent features unique to each LLM. This enables the model to learn more discriminative and robust representations for attribution (\textbf{Phase 2}).

\subsection{Phase 1: Constructing Dataset}
\label{sec:data_construction}
In this phase, we construct \textit{FD-Dataset}. The process involves four main steps: preparing source corpora, constructing seeds and prompts, and querying a wide range of LLMs.

\noindent\textbf{Source Data Preparation.}
We begin by collecting large-scale English and Chinese corpora~\cite{enwords,ChineseFinewebEdu} that span a wide range of domains and registers. All raw data undergo language-specific preprocessing pipelines, which include cleaning, deduplication, and filtering of sensitive content. This ensures the corpora are both broad and representative of real-world language use while minimizing noise and inappropriate content.

\noindent\textbf{Seed and Prompt Construction.}
To construct seeds and generate diverse, controlled prompts (see Algorithm~\ref{alg:seed_prompt} and Figure~\ref{fig:combined}), we employ a two-step process. 
For English, candidate seeds are selected using LLM-based scoring to ensure informativeness and broad coverage. 
For Chinese, seeds are extracted through word segmentation~\cite{jiao2018LAC}, sensitive term filtering, and balancing across linguistic categories. 
Ultimately, we extract over 370,000 English words and more than 75 million unique Chinese words to form a comprehensive multilingual seed pool, which serves as the foundation for constructing content-rich prompts.

\begin{algorithm}[t]
\caption{Seed and Prompt Construction}
\label{alg:seed_prompt}
\begin{algorithmic}[1]
\State \textbf{Input:} English corpus $\mathcal{D}_{\mathrm{en}}$, Chinese corpus $\mathcal{D}_{\mathrm{zh}}$, thresholds $\tau_{\mathrm{en}}, \delta$, target size $N$
\State \textbf{Output:} Final prompt set $\mathcal{P}$

\State $\mathcal{S}_{\mathrm{en}}\gets\varnothing,\;\; \mathcal{S}_{\mathrm{zh}}\gets\varnothing$

\For{$t\in\mathcal{D}_{\mathrm{en}}$}
    \State $s\gets \sigma(t)$ \Comment{$\sigma$: salience \& coverage score}
    \If{$s\ge \tau_{\mathrm{en}}$}
        \State $\mathcal{S}_{\mathrm{en}}\gets\mathcal{S}_{\mathrm{en}}\cup\{t\}$
    \EndIf
\EndFor

\For{$s\in\mathcal{D}_{\mathrm{zh}}$}
    \For{$w\in \omega(s)$} \Comment{$\omega$: word–segmentation}
        \If{$\neg\textsc{Sens}(w)\land \textsc{Bal}(w,\mathcal{S}_{\mathrm{zh}})$} 
            \Comment{$\neg\textsc{Sens}(w)$: $w$ is not sensitive; $\textsc{Bal}$: category balancing}
            \State $\mathcal{S}_{\mathrm{zh}}\gets\mathcal{S}_{\mathrm{zh}}\cup\{w\}$
        \EndIf
    \EndFor
\EndFor

\State $\mathcal{S}\gets\mathcal{S}_{\mathrm{en}}\cup\mathcal{S}_{\mathrm{zh}}$ \Comment{bilingual seed pool}
\State $\mathcal{P}\gets\varnothing$

\While{$|\mathcal{P}|<N$}
    \State $T\gets\mathrm{Sample}(\mathcal{S},3)$
    \If{$\mathrm{Similarity}(T)\le\delta$}
        \If{$\pi(T)$} \Comment{$\pi$: pilot generation success}
            \State $\mathcal{P}\gets\mathcal{P}\cup\{\rho(T)\}$ \Comment{$\rho$: prompt constructor}
        \EndIf
    \EndIf
\EndWhile
\State \Return $\mathcal{P}$
\end{algorithmic}
\end{algorithm}

\begin{table}[t]
\centering
\caption{List of Evaluated Language Models}
\label{tab:evaluated_models}
\begin{tabular}{ccc}
\toprule
\textbf{Category} & \textbf{Model} & \textbf{Version/Parameter} \\
\midrule
\multirow{11}{*}{\textbf{Proprietary}} 
& GPT-4o~\cite{openai2024gpt4technicalreport} & gpt-4o-2024-11-20 \\
& GPT-4o-mini~\cite{openai2024gpt4technicalreport} & gpt-4o-mini-2024-07-18 \\
& GPT-3.5~\cite{GPT-3.5-Turbo} & gpt-3.5-turbo-0125 \\
& Gemini-1.5~\cite{team2024gemini} & gemini-1.5-flash \\
& Claude3.5-haiku~\cite{claude35haiku} & claude-3-haiku-20240307 \\
& Qwen-turbo~\cite{qwen25} & qwen-turbo-1101 \\
& Deepseek~\cite{deepseekai2024deepseekv3technicalreport} & deepseek-v2 \\
& Moonshot~\cite{mootshot} & moonshot-v1 \\
& Doubao~\cite{doubao} & Doubao-lite-32k \\
& Baichuan4~\cite{baichuan4} & Baichuan4-Air \\
& GLM4-Flash~\cite{glm2024chatglm} & glm-4-flash \\
& GLM4-Plus~\cite{glm2024chatglm} & glm-4-plus \\
\hdashline
\multirow{8}{*}{\textbf{Open-Source}} 
& Qwen2.5~\cite{qwen25} & 14B \\
& Llama3.1~\cite{grattafiori2024llama3herdmodels} & 8B \\
& Llama2~\cite{touvron2023llama2openfoundation} & 7B \\
& Gemma2~\cite{gemmateam2024gemma2improvingopen} & 9B \\
& GLM4~\cite{glm2024chatglm} & 9B \\
& InternLM2~\cite{cai2024internlm2} & 7B \\
& Mistral~\cite{jiang2023mistral} & 7B \\
& Yi~\cite{ai2024yi} & 6B \\
\bottomrule
\end{tabular}
\end{table}

\noindent\textbf{LLM Querying.}
The curated prompts are submitted to a wide pool of 20 LLMs (see Table~\ref{tab:evaluated_models}), including both open-source and proprietary series. This represents significantly broader coverage than prior studies~\cite{zeng2023huref,he2024mgtbench,Shi_2024}, and is crucial for learning and evaluating generalizable model fingerprints. Each model is queried under default settings, and the outputs are post-processed with both automated and manual screening to filter out invalid or inappropriate responses.

\begin{table}[t]
\caption{\textit{FD-Dataset} Distribution by Language and Temperature.}
\begin{center}
\label{tab:language_distribution}
\begin{tabular}{ccccccc}
\toprule
\multirow{2}{*}{\textbf{Language}} & 
\multicolumn{5}{c}{\textbf{Temperature}} & 
\multirow{2}{*}{\textbf{Total}} \\
\cmidrule(lr){2-6}
 & \textbf{0} & \textbf{0.3} & \textbf{0.5} & \textbf{0.7} & \textbf{1} &  \\
\midrule
\textbf{en} & 13{,}000 & 3{,}000 & 13{,}000 & 3{,}000 & 13{,}000 & \textbf{45{,}000}  \\ 
\textbf{zh} & 13{,}000 & 3{,}000 & 13{,}000 & 3{,}000 & 13{,}000 & \textbf{45{,}000}  \\ 
\textbf{Total} & \textbf{26{,}000} & \textbf{6{,}000} & \textbf{26{,}000} & \textbf{6{,}000} & \textbf{26{,}000} & \textbf{90{,}000}  \\ 
\bottomrule
\end{tabular}
\end{center}
\end{table}

\noindent \textbf{Dataset Statistics.}
\textit{FD-Dataset} contains 90,000 LLM-generated samples, evenly divided between English and Chinese. Each sample is assigned a generation temperature ($T \in\{0,0.3,0.5,0.7,1\}$) to simulate both deterministic and creative model behaviors, with $T=0.3$ and $T=0.7$ used exclusively for test sets. This setup rigorously assesses model generalization to previously unseen generation regimes.

\begin{figure}[t]
\centering
\includegraphics[width=0.75\linewidth]{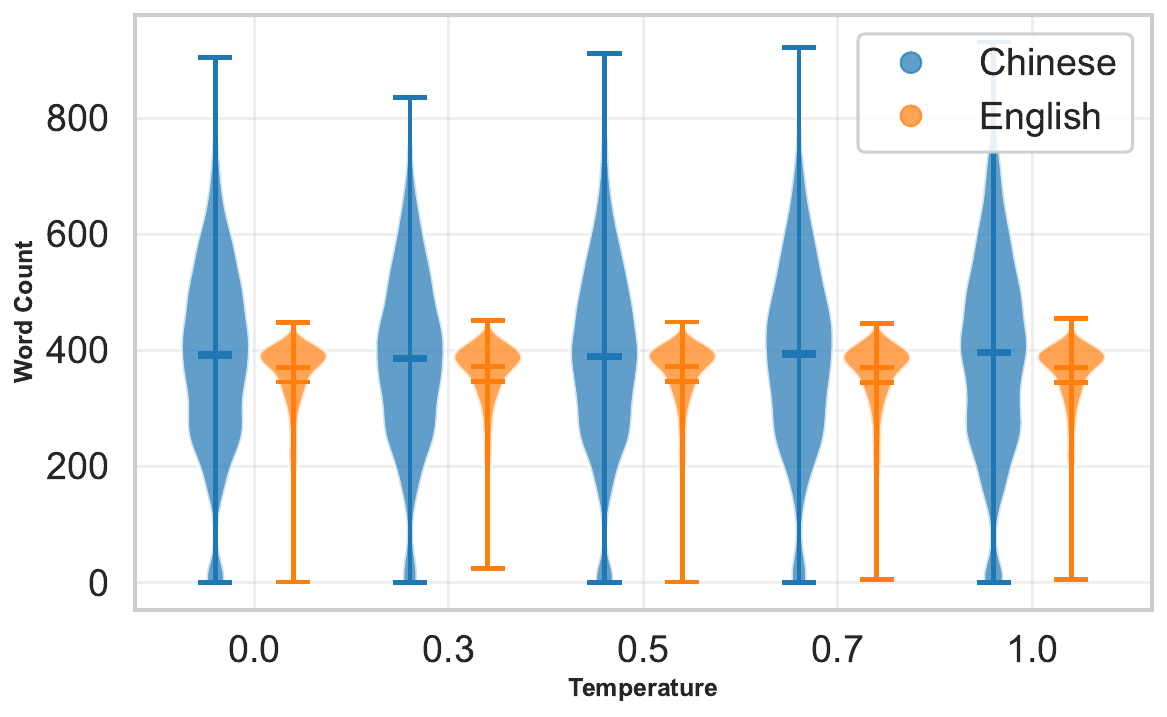}
\caption{Word count distributions for Chinese and English texts generated by LLMs at different temperature settings.}
\label{fig:wordcount}
\end{figure}

\textit{FD-Dataset} feature coverage of both topical and stylistic aspects. For each entry, prompts are generated by randomly combining seed terms, promoting domain and style diversity. As Figure~\ref{fig:wordcount} shows, word count distributions are broad and stable across temperature and language, demonstrating the dataset’s ability to capture both deterministic and creative outputs. English samples generally have higher word counts than Chinese due to linguistic characteristics and tokenization.

\subsection{Phase 2: Learning Fingerprints}
\label{sec:representation_adaptation}

Detecting the unique fingerprints left by different LLMs requires a principled representation learning approach. To better understand and formalize this process, let $\theta$ denote the different LLM. An LLM can be abstracted as:
\begin{equation*}
\mathcal{M}_{\theta} : x \mapsto y \quad \text{where} \quad y \sim P_{\theta}(y \mid x),
\end{equation*}
Given an input $x$, the model $\mathcal{M}_{\theta}$ produces a textual output $y$ sampled from its learned conditional distribution $P_{\theta}(y \mid x)$. Even with the same input, different models tend to diverge in style, content, and reasoning due to differences in $P_{\theta}$. Thus, the fingerprint of a model can be formally defined as a transformation of its generation distribution:
\begin{equation*}
\text{Fingerprint}(\mathcal{M}_{\theta}) := \Phi(P_{\theta}),
\end{equation*}
where $\Phi(\cdot)$ denotes a feature extraction function capturing statistical and stylistic characteristics, such as rare word choices, tokenizer boundary patterns, implicit punctuation habits, and sampling noise. In practice, since $\theta$ and $P_{\theta}$ are unobservable, we analyze a set of outputs $\{y_i\}_{i=1}^n$ to derive a model-unique fingerprint representation:
\begin{equation*}
\mathbf{F}_{\theta} = \Phi(\{y_i\}_{i=1}^n).
\end{equation*}
Based on these fingerprint vectors, a classifier $C$ can be trained to identify the source model:
\begin{equation*}
C(\mathbf{F}_{\theta}) = \hat{\mathcal{M}}, \quad \hat{\mathcal{M}} \in \{\mathcal{M}_1, \ldots, \mathcal{M}_K\}.
\end{equation*}
To move beyond manual feature engineering, we introduce a data-driven, model-based fingerprint extraction strategy using PEFT. As illustrated in Figure~\ref{fig:framework}, our fingerprint learning approach adapts a frozen, pre-trained model with LoRA modules.

In this black-box attribution setting, only the LLMGT is accessible. The corresponding inference prompts used for generation are also available (see Figure~\ref{fig:combined}). Rather than relying on handcrafted features, we leverage the semantic-rich internal representations of a frozen base/backbone model. Specifically, for a given input, we extract the latent feature representation $\mathbf{F} \in \mathbb{R}^d$ from the frozen model. To make these representations more discriminative for attribution, we apply LoRA to selected projection matrices, introducing trainable updates $\Delta\mathbf{W}$ that shift the base/backbone features into attribution-relevant subspaces:
\begin{equation*}
\mathbf{F}' = \mathbf{F} + \Delta\mathbf{F},
\end{equation*}
where $\mathbf{F}'$ denotes the adapted feature after LoRA, capturing the distinctive generation characteristics of each LLM.

To guide the model toward learning attribution-relevant features, we frame the task as a multi-class classification problem, where each class corresponds to a specific LLM fingerprint. We train the model using the standard Cross-Entropy (CE) loss, which encourages the model to assign a high probability to the correct class label for each input. The CE loss for a sample $i$ with true label $y_i$ is defined as:
\begin{equation}
\mathcal{L}_{\text{CE}} = -\frac{1}{N} \sum_{i=1}^{N} \log p(y_i \mid x_i)
\label{eq:ce_loss}
\end{equation}
where $p(y_i \mid x_i)$ denotes the predicted probability for the true class $y_i$.
This loss not only promotes the learning of discriminative representations for attribution but also provides stable and efficient convergence in large-scale scenarios. Compared to more complex alternatives such as contrastive objectives (see Appendix~\ref{appendix:aco}), CE loss offers a straightforward and robust training objective for LLM attribution.

The Last Hidden States block in Figure~\ref{fig:framework} visualizes how LoRA modules guide these latent representation shifts. For example, class centroids move directionally in the latent space, forming more precise decision boundaries.

Finally, the adapted features are passed to a lightweight classifier (\NAME) to predict the source LLM:
\begin{equation}
\hat{y} = \text{softmax}(\mathbf{W}_m \mathbf{F}' + \mathbf{b}_c),
\label{eq:classifier}
\end{equation}
where $\mathbf{W}_m \in \mathbb{R}^{C \times d}$ is the classifier weight matrix that projects the $d$-dimensional features to $C$ class logits (corresponding to the number of LLM sources), and $\mathbf{b}_c \in \mathbb{R}^C$ is an optional bias term.
This design is model-agnostic and can be directly extended to other pre-trained LLMs. We will discuss the effectiveness of different backbones shortly in Section~\ref{sec:rq:performance}.

\section{Threat Model}\label{sec:attack}
\textit{Insight 3} emphasizes the importance of robustness against adversarial edits and distribution shifts. In this section, we evaluate the stability of model fingerprints under three challenging scenarios, including cross-lingual translation, polishing, and Synonym Substitution attacks. To address this, our threat model specifies the following adversarial goals and capabilities:

\textbf{c c} The adversary aims to obscure the true origin of LLMGT, evading models designed to attribute content to its source. Specifically, given a passage produced by a target LLM, the adversary seeks to transform the text in ways that prevent attribution while preserving its semantics and utility.

\textbf{Adversary’s Capability.} We assume the adversary has access to one or more proprietary or open-source LLMs different from the target model being attributed. The adversary does not know the internal structure or parameters of the attribution detector but can query it in a black-box manner (i.e., observe outputs for given inputs). The adversary can use LLMs to perform high-quality transformations on the original text.

\subsection{Attack Scenarios and Methodologies}

\begin{figure}[t]
\centering
\includegraphics[width=\linewidth]{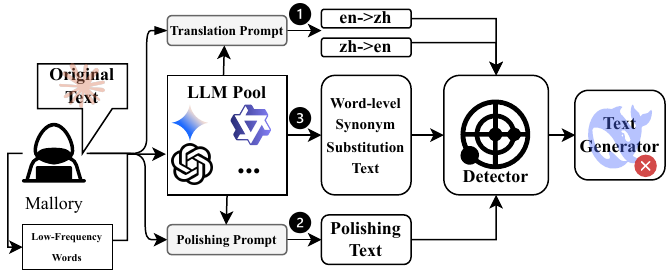}
\caption{The Scenario of Robustness Threat.}
\label{fig:frameattaack}
\end{figure}

To evaluate robustness, we consider three attack scenarios (see Figure~\ref{fig:frameattaack}): 
\ding{182}~LLM-based Translation Attack
\ding{183}~LLM-based Polishing Attack, and 
\ding{184}~Word-Level Synonym Substitution Attack.
These attacks are particularly attractive because they are low-cost and scalable, requiring only access to general LLMs and simple prompts.
\label{subsec:tm}

\noindent \ding{182}~\textbf{LLM-based Translation Attack:}
The adversary perturbs the original text $x$ by translating it into another language using a general LLM, resulting in a transformed text $x^*$.
This approach preserves the semantic content but substantially modifies surface linguistic features, such as word order, synonym usage, and sentence structure, which are often relied upon by attribution models.
Translation introduces paraphrasing effects that undermine the effectiveness of attribution classifiers.
\begin{equation}
\begin{aligned}
x^* = T&ranslation(x) \\
\arg\max_{y_i \in \mathcal{Y}} P(y_i|x^*) &\neq \arg\max_{y_i \in \mathcal{Y}} P(y_i|x)
\end{aligned}
\end{equation}
Here, $x = t_1t_2...t_n$ denotes the original text as a sequence of tokens from the LLM's vocabulary $V$.

\noindent\ding{183}~\textbf{LLM-based Polishing Attack:}
In this scenario, the adversary uses a different LLM to polish the original text $x$, yielding $x^*$ that is semantically equivalent but distinct in style. The polishing process typically normalizes register, removes typos, compresses verbose phrases, and substitutes rare or informal expressions with more standard alternatives. 
Because many attribution detectors rely on shallow stylistic cues such as sentence length, punctuation, or phraseology, these edits can significantly reduce the ability to detect specific fingerprints.
Like the translation attack, this is a black-box method that requires only access to a capable rewriting LLM. 
\begin{equation}
\begin{aligned}
x^* = Po&lishing(x) \\
\arg\max_{y_i \in \mathcal{Y}} P(y_i \mid x^*) &\neq \arg\max_{y_i \in \mathcal{Y}} P(y_i \mid x)
\end{aligned}
\end{equation}

\begin{algorithm}[t]
\caption{Word-Level Synonym Substitution Attack}
\label{alg:synonym}
\begin{algorithmic}[1]
\State \textbf{Input:} Text $x$; stopword set $\mathcal{S}$; number of substitutions $k$; LLM API
\State \textbf{Output:} Perturbed text $x^*$
\State $W \gets \mathrm{Tokenize}(x)$
\State $W' \gets W \setminus \mathcal{S}$
\State $W_{\mathrm{sorted}} \gets \mathrm{Sort}_{\mathrm{asc}}(\, \mathrm{Freq}(W'))$
\State $C \gets \{w_i\}_{i=1}^{k} \subseteq W_{\mathrm{sorted}}$
\For{each $w \in C$}
    \State $w' \gets \mathrm{LLM\_API}(x,\, w)$
    \State $x \gets x[w \mapsto w']$
\EndFor
\State $x^* \gets x$
\State \Return $x^*$
\end{algorithmic}
\end{algorithm}
\noindent\ding{184}~\textbf{Word-Level Synonym Substitution Attack:}
We propose a black-box, model-agnostic lexical attack that perturbs a small number of rare content words in $x$ by replacing them with context-aware synonyms generated by an external LLM, while stopwords and frequent words are left unchanged. This preserves semantic meaning yet disrupts the lexical distribution that attribution models exploit.

The detailed procedure is given in Algorithm~\ref{alg:synonym}. Briefly, we identify rare non-stopword tokens in the input text and use an external LLM to substitute them with suitable synonyms.

\section{Evaluation}
\label{exps}
\subsection{\textbf{Experimental Setup and Research Questions}}\label{expset}
\noindent\textbf{Baseline}. 
To comprehensively evaluate our approach, we compare it against both metric-based and model-based detection methods.

\emph{Metric-based methods} directly compute statistics on the generated text without requiring additional model training. We include six representative approaches: Entropy~\cite{gehrmann2019gltr}, Rank~\cite{gehrmann2019gltr}, GLTR~\cite{gehrmann2019gltr}, Log-Likelihood~\cite{solaiman2019releasestrategiessocialimpacts}, Log-Rank~\cite{mitchell2023detectgpt}, and LPR~\cite{su2023detectllmleveraginglogrank}.

\emph{Model-based methods} require training a separate classifier on labeled data to distinguish between different LLMs. We evaluate five representative methods: DetectGPT~\cite{mitchell2023detectgpt}, ChatGPT-D~\cite{guo2023closechatgpthumanexperts}, OpenAI-D~\cite{solaiman2019releasestrategiessocialimpacts}, LM-D~\cite{ippolito2020automaticdetectiongeneratedtext}, and POGER~\cite{shi2024ten}. Each classification model is trained on the LLMGT corpus following the respective method’s protocol. Notably, POGER is only applied to the English corpus due to the limitations of the model.

\noindent\textbf{Multilingual Evaluation.}
All experiments are conducted on both English (en) and Chinese (zh) datasets to assess the bilingual fingerprinting performance of each method.

\noindent\textbf{Datasets for Robustness and Generalization.}
For the OOD setting, we collected a total of 1,200 samples from QA datasets, evenly split between English and Chinese. For each new LLM not included in the original training set, we provide 50 samples.
For robustness evaluation, we constructed three types of adversarial datasets. For polishing and translation attacks, we generated a total of 1,200 high-quality samples (600 for polishing and 600 for translation), evenly split between English and Chinese, using GPT-4.1-based prompts. 
For the synonym substitution attack, we randomly selected an LLM different from the original generator (see Table~\ref{tab:evaluated_models} and Appendix~\ref{llms}) to provide context-aware synonym replacements. Specifically, we created 8,000 samples by substituting either three or five content words per text, again ensuring a balanced split between English and Chinese.

\noindent \textbf{Other PEFT Methods.}
We further evaluated several other representative PEFT methods, including standard DoRA~\cite{dora2024}, LoRA$+$~\cite{hayou2024lora+}, AdaLoRA~\cite{zhang2023adaptive}, and QLoRA~\cite{dettmers2023qlora}, using their official implementations and recommended hyperparameters. We set the rank of all PEFT methods to be consistent with \NAME.

\noindent\textbf{Hyperparameters.}
For \NAME, we use a batch size of 2 and AdamW optimizer~\cite{loshchilov2017decoupled} with an initial learning rate of $1\mathrm{e}{-4}$; other methods are trained using their recommended default settings.

\noindent \textbf{Metrics.}
We employ a comprehensive set of evaluation metrics to assess both classification performance and adversarial robustness. For model performance, we report \textit{Accuracy (Acc)}, \textit{Macro Precision (MacP)}, \textit{Macro Recall (MacR)}, and \textit{Macro F1 (MacF1)}. 

To evaluate adversarial robustness, we use several criteria. \textit{Attack Success Rate (ASR)} measures the proportion of adversarial examples that are misattributed to the wrong source model. \textit{Text Similarity Rate (TSR)} is calculated as the cosine similarity between the embeddings of adversarial and original texts, reflecting the preservation of meaning. \textit{Perplexity (PPL)} indicates the fluency of generated text, with lower values denoting higher naturalness. Additionally, we include \textit{COMETKiwi}~\cite{rei2022cometkiwi}, a reference-free metric for translation quality estimation, which produces sentence-level quality scores and word-level acceptability tags.

\NAME is evaluated based on the following research questions: 
\begin{enumerate}[label=\textbf{RQ\arabic*:}, leftmargin=*]
    \item \label{rq:performance} Is the performance of \NAME improved compared to other baseline methods on \textit{FD-Dataset}?
    \item \label{rq:lora} Why is LoRA effective in this task?
    \item \label{rq:generalization} What is the generalization capability of \NAME across unseen models and domains?
    \item \label{rq:robustness} How robust is \NAME against adversarial attacks?
    \item \label{rq:influence} How do training set size, temperature settings, and LoRA parameters influence the detection accuracy of \NAME?
\end{enumerate}

\subsection{\ref{rq:performance} Performance Improvement}
\label{sec:rq:performance}

\begin{table}[t]
    \centering
    \caption{Comparison of the average prediction metrics for LLMGT under three generation-temperature settings ($T_{\text{gen}}\!\in\!\{0,0.5,1\}$).  \textbf{Prompt} denotes that \NAME is trained only with prompt tuning; \textbf{Q} denotes the Qwen2.5-Instruct-7B model.}
    \label{tab:baselines}
    \setlength{\tabcolsep}{5pt}
    \begin{tabular}{ccccc}
        \toprule
        \textbf{Method} & \textbf{Acc(\%)} & \textbf{MacP(\%)} & \textbf{MacR(\%)} & \textbf{MacF1(\%)} \\\midrule
        Entropy        &  6.36 &  4.37 &  6.36 &  2.62 \\
        Rank           &  6.46 &  4.87 &  6.46 &  2.78 \\
        DetectGPT      &  9.07 &  5.81 &  9.07 &  3.12 \\
        LRR            &  9.78 &  6.75 &  9.78 &  6.50 \\
        GLTR           & 11.16 &  6.72 & 11.16 &  6.70 \\
        Log-Likelihood & 10.42 &  7.62 & 10.42 &  7.54 \\
        Log-Rank       & 11.20 &  8.83 & 11.20 &  8.32 \\
        POGER     & 44.48 & 35.84 & 44.08 & 35.34 \\
        ChatGPT-D      & 46.28 & 46.68 & 46.28 & 44.52 \\  
        OpenAI-D       & 73.84 & 75.88 & 73.84 & 73.95 \\ 
        LM-D           & 74.58 & 75.77 & 74.58 & 74.48 \\ 
        FDLLM (Prompt) & 90.18 & 92.97 & 90.18 & 90.89 \\ 
        \textbf{FDLLM (Q)} & \textbf{96.56} & \textbf{96.57} & \textbf{96.56} & \textbf{96.56} \\ \bottomrule
    \end{tabular}
\end{table}

Table~\ref{tab:baselines} presents an aggregated comparison of eleven baselines and our proposed \NAME.  
Several observations can be drawn from the results.

\textbf{Traditional heuristics are of limited effectiveness.}  
Metrics such as Entropy, Rank, and DetectGPT yield Macro F1 scores below~4\,\%, suggesting that handcrafted statistics alone are insufficient for distinguishing the fine-grained classes present in our dataset.

\textbf{Intermediate methods provide only marginal improvements.}  
Approaches like GLTR and Log-Likelihood achieve Macro F1 scores of approximately~7–8\,\%, but still fall short of practical requirements. This highlights the limitations of shallow feature engineering and standard likelihood-based scoring. In contrast, fine-tuning emerges as essential for achieving competitive accuracy.

Supervised baselines such as ChatGPT-D, POGER, and OpenAI-D further improve Macro F1 scores to the 44–74\,\% range. However, our method consistently outperforms these approaches.  
Specifically, \NAME(Q) achieves a Macro F1 of \textbf{96.56\,\%}, representing a relative gain of 29 points over the strongest baseline (OpenAI-D). These results demonstrate that targeted PEFT can offer substantial gains in balancing generation stability and classification accuracy for LLM attribution.

\begin{figure}[t]
\centering 
\includegraphics[width=\linewidth]{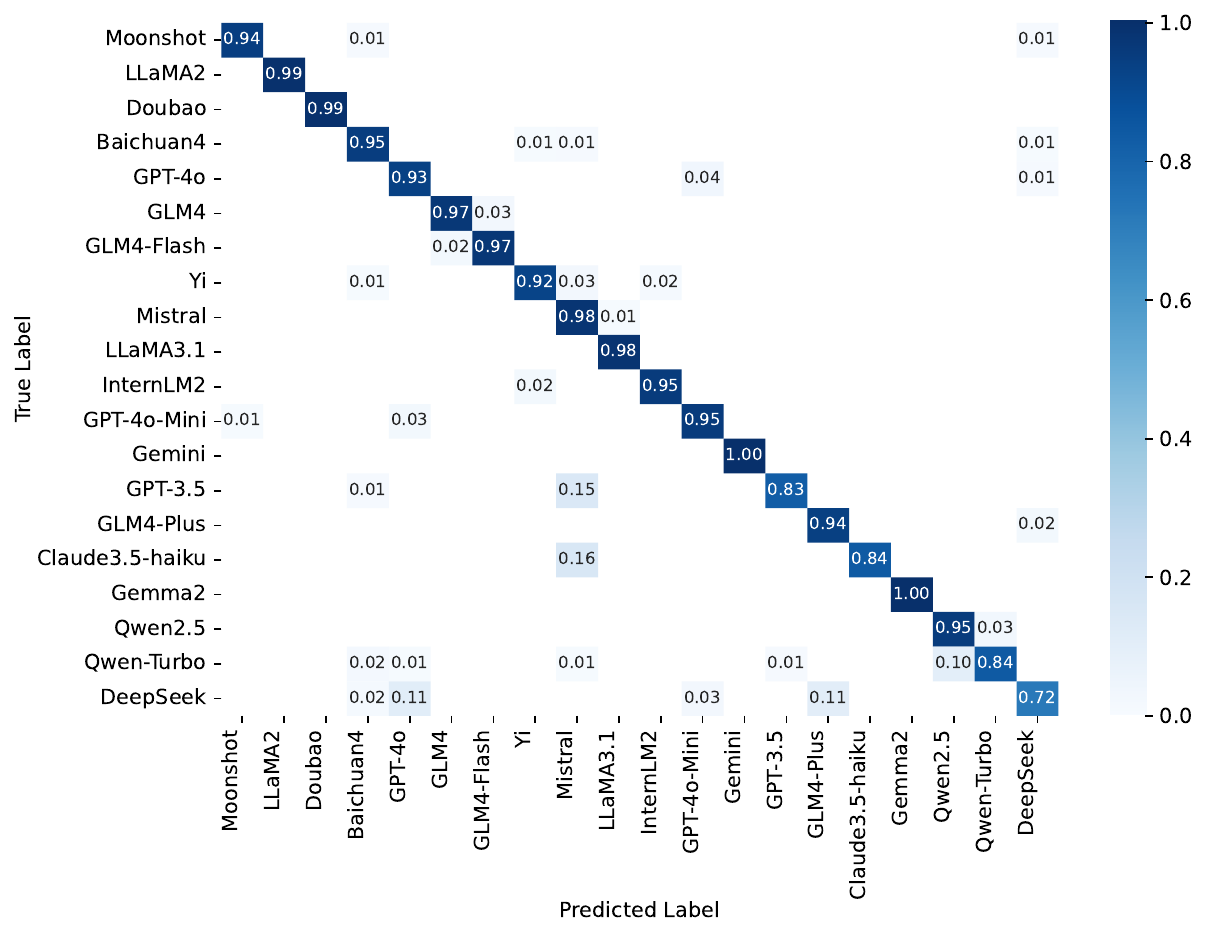}
\caption{Confusion matrix illustrating the classification performance of \NAME. Values less than 0.01 are not shown for better visualization.}
\label{fig4}
\end{figure}

\begin{table}[t]
    \centering
    \caption{Results obtained from training with different base/backbone models. \textbf{Type} denotes the model family or training objective: “Instruct” for instruction-tuned, “Reasoning” for models focused on reasoning tasks, and “Distill” for distilled models.}
    \label{tab:model_performance}
    \setlength{\tabcolsep}{5pt}
    \begin{tabular}{ccccc}
        \toprule
        \textbf{Model} & \textbf{Parameter} & \textbf{Type} & \textbf{Acc(\%)} & \textbf{MacF1(\%)} \\ \midrule
        \textbf{Qwen2.5}      & \textbf{7B}   & \textbf{Instruct}      & \textbf{96.56} & \textbf{96.56} \\
        Qwen3                & $0.6B$          & Reasoning             & 88.39 & 88.38 \\
        Qwen3                & $1.7B$          & Reasoning             & 91.24 & 91.23 \\
        Qwen3                & $4B$            & Reasoning             & 93.23 & 93.22 \\
        Qwen3                & $8B$            & Reasoning             & 94.62 & 94.62 \\
        DeepSeek             & $7B$            & Instruct              & 94.16 & 94.16 \\
        DeepSeek-R1          & $7B$            & Distill-Qwen          & 92.67 & 92.66 \\
        DeepSeek-R1          & $8B$            & Distill-Llama         & 94.05 & 94.04 \\
        GLM4-0414            & $9B$            & Instruct              & 95.12 & 95.12 \\
        GLM-Z1-0414          & $9B$            & Reasoning             & 94.70 & 94.70 \\ 
        MiMo                 & $7B$            & Instruct              & 94.67 & 94.67 \\
        MiMo                 & $7B$            & Reasoning             & 93.93 & 93.93 \\ \bottomrule
    \end{tabular}
\end{table}
Figure~\ref{fig4} illustrates the confusion matrix \NAME tends to confuse GLM4 and GLM4 Flash, likely due to their high similarity. A similar issue occurs between Qwen2.5 and Qwen Turbo. \NAME still has room for improvement. For instance, when handling the Deepseek model, \NAME frequently misclassifies it as other models, such as GPT4o and GLM4 Plus.

Table~\ref{tab:model_performance} further breaks down the attribution performance of \NAME across different backbone models, enabling a detailed analysis of backbone effectiveness. Although the Qwen2.5-7 B-Instruct model attains the highest Macro F1 score (96.56\%), outperforming several larger reasoning-oriented variants such as Qwen3~\cite{qwen3}, GLM4-0414~\cite{glm2024chatglm}, and MiMo~\cite{xiaomi2025mimo}, the overall differences among various backbones remain limited. This suggests that, for the \NAME framework, the choice of backbone has only a modest impact on final attribution accuracy. These results indicate that even lightweight, instruction-tuned models can achieve competitive performance.

\begin{tcolorbox}
\textbf{Take-aways:}
PEFT on instruction-tuned models enables \NAME to achieve a Macro F1 score of 96.6\%, outperforming the strongest baseline by 22 points. Importantly, the final attribution performance remains similar across different backbone models, highlighting the practicality and flexibility of our approach.
\end{tcolorbox}

\begin{figure*}[ht]
  \centering
  \begin{subfigure}[t]{0.49\linewidth}
    \centering
    \includegraphics[width=\linewidth]{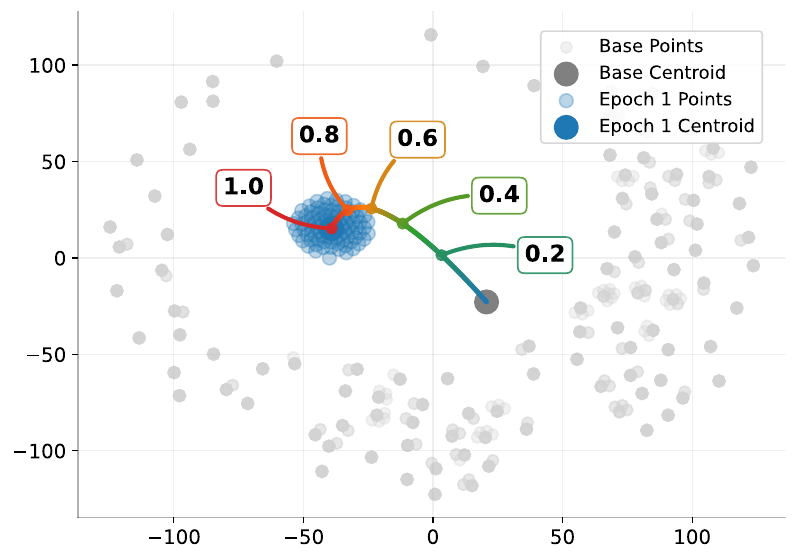}
    \caption{Baichuan4}
  \end{subfigure}%
  \hfill
  \begin{subfigure}[t]{0.49\linewidth}
    \centering
    \includegraphics[width=\linewidth]{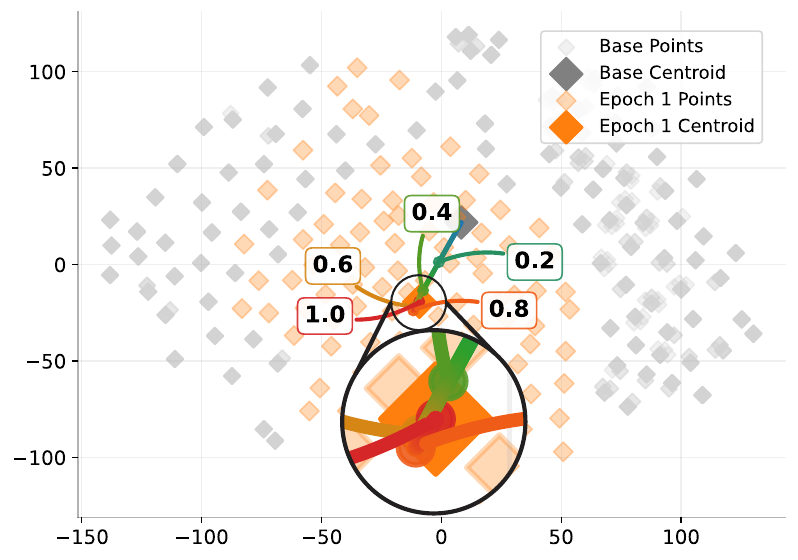}
    \caption{DeepSeek}
  \end{subfigure}
  \caption{In each panel, light-gray markers show the original embeddings produced by the frozen base/backbone model (\textit{Base Points}); the dark-gray symbol (circle in a diamond in b) marks their centroid (\textit{Base Centroid}).  
  The colored poly-line traces the low-rank displacement of the centroid as training proceeds from the start (\(0\)) to one full epoch (\(1\)); numeric call-outs (0.2–1.0) indicate \textit{the fraction of the epoch completed}.  
  Blue (a) / Yellow (b) markers depict the embedding distribution at the end of the first epoch (\textit{Epoch 1 Points}), while the solid dark-blue / dark-yellow symbol marks the corresponding updated centroid (\textit{Epoch 1 centroid}).  
  In both cases, a single epoch of LoRA tuning smoothly steers the centroid into the new cluster and noticeably tightens intra-class dispersion, illustrating how a low-rank update rapidly enhances linear separability.}
  \label{fig:lora_trajectories}
\end{figure*}
\subsection{\ref{rq:lora} LoRA Effectiveness Analysis} 
\label{sec:rq2}

LoRA-based adaptation enables robust model attribution by efficiently structuring the feature space to cluster embeddings from the same model while separating those from different sources (see Figure~\ref{fig:oc}).
Figure~\ref{fig:lora_trajectories} provides a visual comparison of LoRA’s effect on two representative LLMs:

\textbf{Class Centroid Translation:} For each model, the centroid of its feature distribution (dark gray) is progressively steered toward a new, well-separated location, as indicated by the trajectory line that transitions from blue (training start) to red (epoch end). The numeric call-outs (0.2–1.0) mark the fraction of the epoch completed, reflecting the dynamics of centroid migration during LoRA adaptation.

\textbf{Intra-class Compactness and Inter-class Separation:} A direct comparison of panels (a) and (b) highlights the varying effectiveness of LoRA adaptation. In Figure~\ref{fig:lora_trajectories}(a) for Baichuan4, the adapted embeddings at epoch 1 (blue points) are tightly clustered around the new centroid (dark blue), resulting in clear intra-class compactness and distinct separation from other clusters. Conversely, in Figure~\ref{fig:lora_trajectories}(b) for DeepSeek, the epoch 1 embeddings (yellow points) are more dispersed, and the updated centroid (dark yellow) is less clearly isolated. This contrast explains the attribution accuracy gap observed in Figure~\ref{fig4}: \NAME performs well for Baichuan4 but is less effective for DeepSeek and similar models. Thus, the clustering structure in the embedding space visually mirrors downstream attribution performance.

These effects are further evidenced by the observed movement of class centroids and the changes in intra-class and inter-class relationships, which can be formally characterized using the definitions in Section~\ref{sec:representation_adaptation} and Equation~\eqref{eq:contrastive_lora}. Specifically, LoRA adaptation leads to a visible shift of the centroid ($\mathbf{c}_0 \rightarrow \mathbf{c}_1$), a qualitative reduction in intra-class dispersion, and improved separation between different model clusters.

Finally, although T-SNE~\cite{van2008visualizing} axes lack explicit geometric meaning, our visualizations consistently show that LLMs from the same organization or model family tend to cluster together. This suggests that attribution signals may reflect not only individual model fingerprints but also shared stylistic or architectural features, such as pretraining data, tokenizer, or fine-tuning paradigms. Such family-level clustering highlights the broader potential for attribution techniques to capture both instance-level and lineage-level structure in LLM outputs, enriching our understanding of “model lineage” within the attribution landscape.
\begin{tcolorbox}
\textbf{Take-aways:}
This analysis offers novel evidence that LoRA adaptation meaningfully restructures the embedding space in LLM attribution. Our findings support the suitability of LoRA for this task and lay the groundwork for more detailed studies of model fingerprinting and attribution in the future.
\end{tcolorbox}

\subsection{\ref{rq:generalization} Generalization of \NAME to Unseen Models and Domains}
\label{sec:RQ3_generalization}

Although \NAME covers the majority of mainstream LLMs, it is crucial to assess its detection capability on previously unseen LLMs. To simulate this scenario, we perform continual training on the original \NAME weights using LoRA with samples from newly introduced models. 

Specifically, we reserve an additional classification head during initial training to facilitate the seamless incorporation of new model classes in subsequent updates. As observed, models belonging to the same family tend to exhibit very similar behaviors, making generalization easier for \NAME. To evaluate this, we design two settings:
(1) New models from already covered families (e.g., GPT-4.1, GPT-4.1-Mini), and
(2) New models from entirely novel families (e.g., Granite3.3, Phi4). 

\begin{table}[t]
\centering
\setlength{\tabcolsep}{4pt}
\caption{Performance of \NAME after adapting to newly introduced models via LoRA. }
\begin{tabular}{cccccc}
\toprule
\multirow{2}{*}{\textbf{LLM}} & \multirow{2}{*}{\textbf{Vendor}} & \multicolumn{3}{c}{\textbf{Accuracy (\%)}} & \multirow{2}{*}{\textbf{MacF1 (\%)}} \\ 
\cmidrule(lr){3-5}
 & & \textbf{en} & \textbf{zh} & \textbf{Average} & \\ \hline
GPT-4.1~\cite{openai41} & OpenAI & 95.00 & 100.00 & 97.50 & 85.23 \\
GPT-4.1-Mini & OpenAI & 100.00 & 90.00 & 95.00 & 84.63 \\ \hdashline
Granite3.3~\cite{granite33} & IBM & 100.00 & 95.00 & 97.50 & 85.18 \\
Phi4~\cite{phi4} & Microsoft & 95.00 & 90.00 & 92.50 & 85.58 \\ \hline
\textbf{ALL Models} & -- & \textbf{97.50} & \textbf{93.75} & \textbf{95.63} & \textbf{85.16} \\ \bottomrule
\end{tabular}
\label{tab:new_model_performance}
\end{table}
Table~\ref{tab:new_model_performance} reports the results after LoRA adaptation. 
For new models within known families, \NAME achieves high detection accuracy, as seen in GPT-4.1, with an accuracy of 95.00\% (en) and 100.00\% (zh), and in GPT-4.1-Mini, with accuracies of 100.00\% (en) and 90.00\% (zh). Similarly strong results are observed for Granite3.3 (100.00\%/95.00\%) and Phi4 (95.00\%/90.00\%), with an overall average accuracy of 95.63\% and a Macro F1 score of 85.16\%. While performance remains robust across all these settings, a slight decrease is observed for models from previously unseen families, reflecting the increased difficulty of generalizing to novel architectures. Notably, the incremental LoRA adaptation for each new model can be completed in under 20 minutes on a single RTX 3090 GPU.

To further assess generalization under real-world distribution shifts, we evaluate \NAME under OOD conditions using a QA-based dataset constructed from Quora~\cite{fan2019eli5}. This setting introduces unique challenges: all questions are in English, with Chinese versions generated by LLM-based translation, leading to translation features. Additionally, popular benchmark questions often have highly similar answers across models, making attribution even more difficult.

As shown in Table~\ref{tab:ood_results}, \NAME achieves the highest performance among all compared black-box detection methods, with an accuracy of 50.25\% and a Macro F1 score of 49.86\%. In contrast, the best method (POGER) achieves only 44.48\% accuracy and 35.34\% Macro F1, while most baselines remain below 20\%. Random guessing would yield an expected accuracy of only 5\%.
To further test cross-domain generalization, we trained \NAME only on the OOD QA-based dataset and evaluated it on the \textit{FD-Dataset}. The Macro F1 score dropped to 19.43\%, suggesting a notable gap between domains and indicating that strong performance in one setting may not directly transfer to another. This result highlights the potential value of utilizing diverse and representative training data.

Error analysis reveals two main challenges: (1) new models from an existing family are easily confused with other family members due to similar generation patterns; (2) open-source models from new families are often confused with other open-source models. These findings underscore both the complexity of the attribution task and the effectiveness of our approach.

\begin{table}[t]
\centering
\caption{Performance of different detection methods under OOD scenarios.}
\setlength{\tabcolsep}{5pt}
\begin{tabular}{ccccc}
\toprule
\textbf{Method} & \textbf{Acc (\%)} & \textbf{MacP (\%)} & \textbf{MacR (\%)} & \textbf{MacF1 (\%)} \\
\midrule
Entropy        & 5.40   & 3.36   & 5.40   & 2.18 \\
Rank           & 5.20   & 5.22   & 5.20   & 2.67 \\
DetectGPT      & 6.80   & 4.04   & 6.80   & 3.31 \\
GLTR           & 8.10   & 6.06   & 8.10   & 3.86 \\
Log-Likelihood & 7.55   & 6.01   & 7.55   & 3.98 \\
LRR            & 8.00   & 5.97   & 8.00   & 4.62 \\
Log-Rank       & 8.70   & 5.78   & 8.70   & 4.70 \\
LM-D           & 14.20  & 17.74  & 14.20  & 9.25 \\
ChatGPT-D      & 14.75  & 17.36  & 14.75  & 14.56 \\
OpenAI-D       & 20.80  & 30.01  & 20.80  & 17.47 \\
POGER         & 44.48  & 35.84  & 44.08  & 35.34 \\
\textbf{FDLLM} & \textbf{50.25}  & \textbf{57.36}  & \textbf{50.25}  & \textbf{49.86} \\
\bottomrule
\end{tabular}
\label{tab:ood_results}
\end{table}

\begin{tcolorbox}
\textbf{Take-aways:}
\NAME demonstrates strong generalization, achieving high accuracy when adapting to previously unseen models with minimal data and rapid incremental updates. The method also retains its advantages in challenging OOD scenarios. These results suggest that our approach can remain effective when applied to new LLMs and domains, providing a practical solution for real-world attribution where continuous model updates and domain shifts are inevitable.
\end{tcolorbox}

\subsection{\ref{rq:robustness} The Robustness of \NAME to Adversarial Attacks}

\begin{table}[ht]
\centering
\caption{Evaluation of Adversarial Attacks. In this table, higher metric values indicate greater attack success or effectiveness.}
\label{table:adv_attack_results}
\setlength{\tabcolsep}{3pt}
\begin{tabular}{cccccc}
\toprule
\multirow{2}{*}{\textbf{Attack}} & 
\multirow{2}{*}{\textbf{Language}} & 
\multirow{2}{*}{\textbf{TSR (\%)}} & 
\multicolumn{3}{c}{\textbf{ASR (\%)}} \\
\cmidrule(lr){4-6}
 &  &  & \textbf{LM-D} & \textbf{OpenAI-D} & \textbf{FDLLM} \\
\midrule
\multirow{2}{*}{Sub(3)} & en & 99.70 & 15.81 & 8.80 & \textbf{3.05} \\
                        & zh & 99.39 & 47.47 & 41.56 & \textbf{5.03} \\
\hdashline
\multirow{2}{*}{Sub(5)} & en & 99.58 & 15.75 & 8.59 & \textbf{3.31} \\
                        & zh & 98.32 & 48.89 & 42.56 & \textbf{6.55} \\
\hdashline
\multirow{2}{*}{Polishing} & en & 93.27 & 39.11 & 33.11 & \textbf{29.67} \\
                        & zh & 94.82 & 49.98 & 53.67 & \textbf{25.29} \\
\hdashline
\multirow{2}{*}{Translation}  & en & 66.11 & 86.56 & 84.37 & \textbf{66.55} \\
                        & zh & 71.94 & 90.11 & 84.79 & \textbf{52.05} \\
\midrule
\textbf{Average} & - & - & \textbf{49.21} & \textbf{44.68} & \textbf{23.94} \\
\bottomrule
\end{tabular}
\end{table}

\begin{figure}[t]
    \centering
    \begin{subfigure}[b]{0.49\columnwidth}
        \centering
        \includegraphics[width=\textwidth]{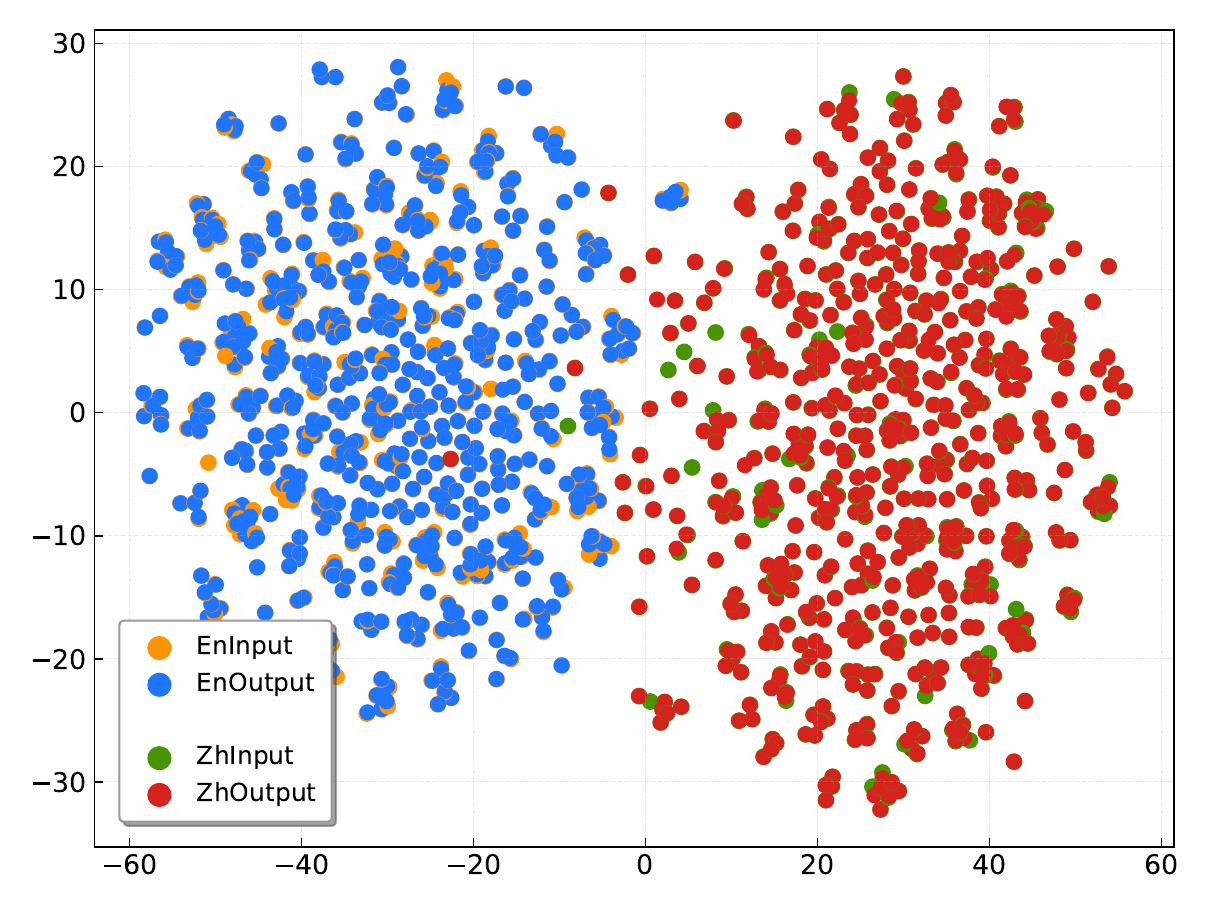}
        \caption{Polishing Attack}
        \label{figPolishing}
    \end{subfigure}
    \begin{subfigure}[b]{0.49\columnwidth}
        \centering
        \includegraphics[width=\textwidth]{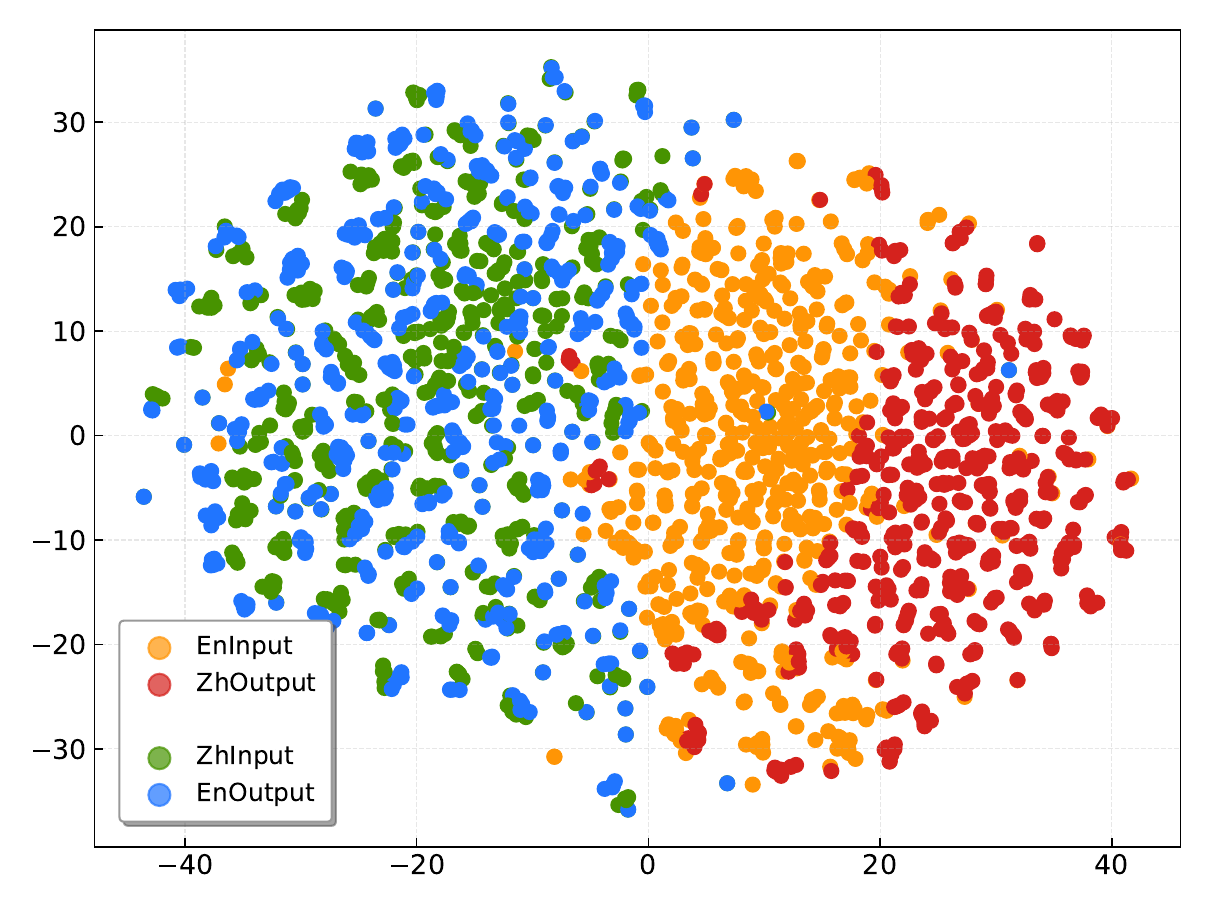}
        \caption{Translation Attack}
        \label{figtrans}
    \end{subfigure}
    \caption{T-SNE visualization of LLM-generated text.}
    \label{fig:combinedembed}
\end{figure}

 The effectiveness of these adversarial modifications is then assessed against three leading model attribution detectors: the top-performing OpenAI-D and LM-D baselines (as identified in Section~\ref{exps}) and our proposed \NAME.

\begin{table}[t]
\centering
\caption{Average performance degradation under Polishing attacks. The \textbf{$\Delta$F1} column indicates the F1 change (\%) compared to the original model performance. The \textbf{$\Delta$Len} column reports the average token count difference between original and adversarial outputs, where positive values indicate that the adversarial outputs are longer.}
\label{table:polishing_attack_results}
\begin{tabular}{lcccc}
\toprule
\multirow{2}{*}{\textbf{Model}} & \multicolumn{3}{c}{\textbf{$\Delta$Len}} & \multirow{2}{*}{\textbf{$\Delta$F1(\%)}} \\
\cmidrule(lr){2-4}
& \textbf{en} & \textbf{zh} & \textbf{Average} & \\ 
\midrule
Baichuan4 & -60.83 & -20.27 & -40.55 & -37.70 \\
Claude3.5-haiku & -13.23 & -3.70 & -8.47 & -43.90 \\
DeepSeek & -58.80 & -11.73 & -35.27 & -40.58 \\
Doubao & -29.63 & -9.70 & -19.67 & -37.78 \\
GLM4 & -28.37 & -20.77 & -24.57 & -19.31 \\
GLM4-Flash & -44.27 & -39.13 & -41.70 & -27.46 \\
GLM4-Plus & -33.30 & -7.60 & -20.45 & -30.25 \\
GPT-3.5 & -116.87 & -2.97 & -59.92 & -43.11 \\
GPT-4o & -76.00 & -19.40 & -47.70 & -38.30 \\
GPT-4o-Mini & -40.90 & -5.17 & -23.03 & -15.95 \\
Gemini & -100.03 & -35.53 & -67.78 & -27.27 \\
Gemma2 & -88.67 & -16.37 & -52.52 & -9.09 \\
InternLM2 & -9.57 & -13.17 & -11.37 & -9.68 \\
LLaMA2 & -19.80 & -28.63 & -24.22 & -18.39 \\
LLaMA3.1 & -105.30 & -10.77 & -58.03 & -11.15 \\
Mistral & -74.07 & -2.27 & -38.17 & -33.78 \\
Moonshot & -58.03 & -4.83 & -31.43 & -17.89 \\
Qwen-Turbo & -72.30 & -36.27 & -54.28 & -17.58 \\
Qwen2.5 & -38.57 & -7.80 & -23.18 & -28.03 \\
Yi & -73.60 & -32.07 & -52.83 & -17.36 \\ \hline
\textbf{Average} & \textbf{-57.11} & \textbf{-16.41} & \textbf{-36.76} & \textbf{-27.27} \\
\bottomrule
\end{tabular}
\end{table}

\noindent \textbf{\ding{182} LLM-based Polishing Attack Results.}
Table~\ref{table:polishing_attack_results} presents the performance degradation under polishing attacks. As illustrated by the T-SNE visualization, polishing attacks induce only modest changes in the feature space: polished texts largely overlap with the originals, remaining within the same language clusters. Correspondingly, the results show that while polishing leads to an average 36.76\% reduction in text length, the average F1 degradation is moderate at 27.27\%. Table~\ref{table:adv_attack_results} presents the ASR for \NAME under a Polishing attack as 29.67\% (en) and 25.29\% (zh). 
This suggests that \NAME captures deeper, persistent features. Polishing alters surface properties, such as token length, but does not eliminate these deeper signals. As a result, \NAME can still identify the source LLM after polishing.

\noindent \textbf{\ding{183} LLM-based Translation Attack Results.}
Table~\ref{table:translation_attack_results} and Figure~\ref{fig:combinedembed} show that LLMGT detectors are considerably more vulnerable to translation attacks, especially at the sentence level. Translation attacks cause a clear and substantial shift in feature space, as evidenced by the separation in T-SNE plots and the quantitative results. 
The T-SNE visualizations reveal that Chinese-to-English translation preserves the original semantic distribution better than English-to-Chinese translation. In the former case, translated embeddings remain closer to the original cluster, while in the latter, a greater separation is observed, indicating higher information loss or alteration.

On average, translation attacks reduce F1 by 54.53\%, nearly doubling the impact of polishing. Notably, the ASR for \NAME under translation attacks rises to 66.55\% (en) and 52.05\% (zh). Even though COMETKiwi scores remain high (average $\sim$0.68), confirming that semantic content is retained, the increase in perplexity ($\Delta$PPL) and the sharp drop in F1 demonstrate that cross-lingual transformation fundamentally disrupts the learned attribution features.

\begin{table}[t]
\centering
\caption{Average performance degradation under Translation attacks.}
\label{table:translation_attack_results}
\begin{tabular}{lccccc}
\toprule
\multirow{2}{*}{\textbf{Model}} 
  & \multicolumn{2}{c}{\textbf{COMETKiwi}} 
  & \multicolumn{2}{c}{\textbf{$\Delta$PPL(\%)}} 
  & \multirow{2}{*}{\textbf{$\Delta$F1(\%)}} \\
\cmidrule(lr){2-3}\cmidrule(lr){4-5}
  & \multicolumn{1}{c}{\textbf{en}} 
  & \multicolumn{1}{c}{\textbf{zh}}
  & \multicolumn{1}{c}{\textbf{en}} 
  & \multicolumn{1}{c}{\textbf{zh}} 
  & \\
\midrule
Baichuan4 & 0.6679 & 0.6470 & 9.30 & 2.11 & -67.76 \\
Claude3.5-haiku & 0.7305 & 0.7695 & 10.13 & 0.43 & -29.00 \\
DeepSeek & 0.6547 & 0.7420 & 16.33 & 2.65 & -90.18 \\
Doubao & 0.7862 & 0.7500 & 11.41 & 5.07 & -60.00 \\
Gemini & 0.6176 & 0.7371 & 14.28 & 2.91 & -77.78 \\
Gemma2 & 0.6799 & 0.7664 & 8.72 & 3.54 & -42.86 \\
GLM4 & 0.6779 & 0.6185 & 8.14 & 2.14 & -56.94 \\
GLM4-Flash & 0.6789 & 0.6166 & 9.03 & 2.29 & -36.07 \\
GLM4-Plus & 0.6515 & 0.5508 & 12.54 & 3.04 & -84.42 \\
GPT-3.5 & 0.7282 & 0.7576 & 8.39 & -3.04 & -49.09 \\
GPT-4o & 0.6523 & 0.6348 & 9.67 & -0.11 & -37.30 \\
GPT-4o-Mini & 0.6504 & 0.6808 & 10.54 & 0.03 & -51.01 \\
InternLM2 & 0.6528 & 0.5955 & 9.56 & 4.32 & -42.09 \\
LLaMA2 & 0.7167 & 0.7385 & 6.53 & 0.40 & -33.08 \\
LLaMA3.1 & 0.6925 & 0.7473 & 7.36 & -0.96 & -38.69 \\
Mistral & 0.6878 & 0.7601 & 11.03 & -2.41 & -67.89 \\
Moonshot & 0.7193 & 0.6281 & 5.44 & 2.16 & -40.72 \\
Qwen-Turbo & 0.6449 & 0.6737 & 11.21 & 3.24 & -64.89 \\
Qwen2.5 & 0.6498 & 0.6636 & 11.64 & 4.58 & -55.15 \\
Yi & 0.6663 & 0.6377 & 8.70 & 2.23 & -48.66 \\
\hline
\textbf{Average} & \textbf{0.6803} & \textbf{0.6858} & \textbf{10.00} & \textbf{1.46} & \textbf{-54.53} \\
\bottomrule
\end{tabular}
\end{table}

\noindent \textbf{\ding{184} Word-Level Synonym Substitution Attack Results.}
\label{sec:results_synonym}
We assess the effectiveness of our proposed word-level synonym substitution attack (denoted as \textbf{Sub($k$)}, where $k$ is the number of substituted words) in both English and Chinese contexts. As reported in Table~\ref{table:adv_attack_results}, the attack achieves exceptionally high semantic preservation, with the average TSR exceeding 98\% in all cases. For example, under the \textbf{Sub(3)} setting, TSR reaches 99.70\% for English and 99.39\% for Chinese, demonstrating that the attack introduces minimal semantic drift.

Despite its implicit nature, this perturbation strategy results in substantial performance degradation in most baseline detectors, which often rely on superficial lexical patterns, such as rare or stylistically distinctive tokens. In contrast, \NAME exhibits remarkable resilience. For instance, in the English \textbf{Sub(3)} scenario, \NAME records an ASR of only 3.05\%, dramatically outperforming LM-D (15.81\%) and OpenAI-D (8.80\%). The gap is even more pronounced in Chinese, where \NAME achieves an ASR of just 5.03\%, compared to 47.47\% for LM-D and 41.56\% for OpenAI-D.

These results highlight a key advantage of \NAME: instead of depending on isolated lexical features such as rare tokens, it captures more holistic generation patterns that embody the underlying stylistic and structural characteristics unique to each LLM.
As a result, \NAME remains robust even under more aggressive perturbations like \textbf{Sub(5)}, where other methods show severe performance drops. These findings provide a promising direction for future research on robust model fingerprinting.

\begin{tcolorbox}
\textbf{Take-aways:}
\NAME exhibits strong robustness against a range of adversarial attacks and consistently outperforms existing baselines. This demonstrates its practical potential for real-world attribution scenarios, where resilience to adversarial modifications is crucial. 
\end{tcolorbox}

\subsection{\ref{rq:influence} Impact of Hyperparameters on \NAME Performance}
\begin{figure*}[ht]
\centering
\begin{minipage}[b]{0.24\textwidth}
    \includegraphics[width=\textwidth]{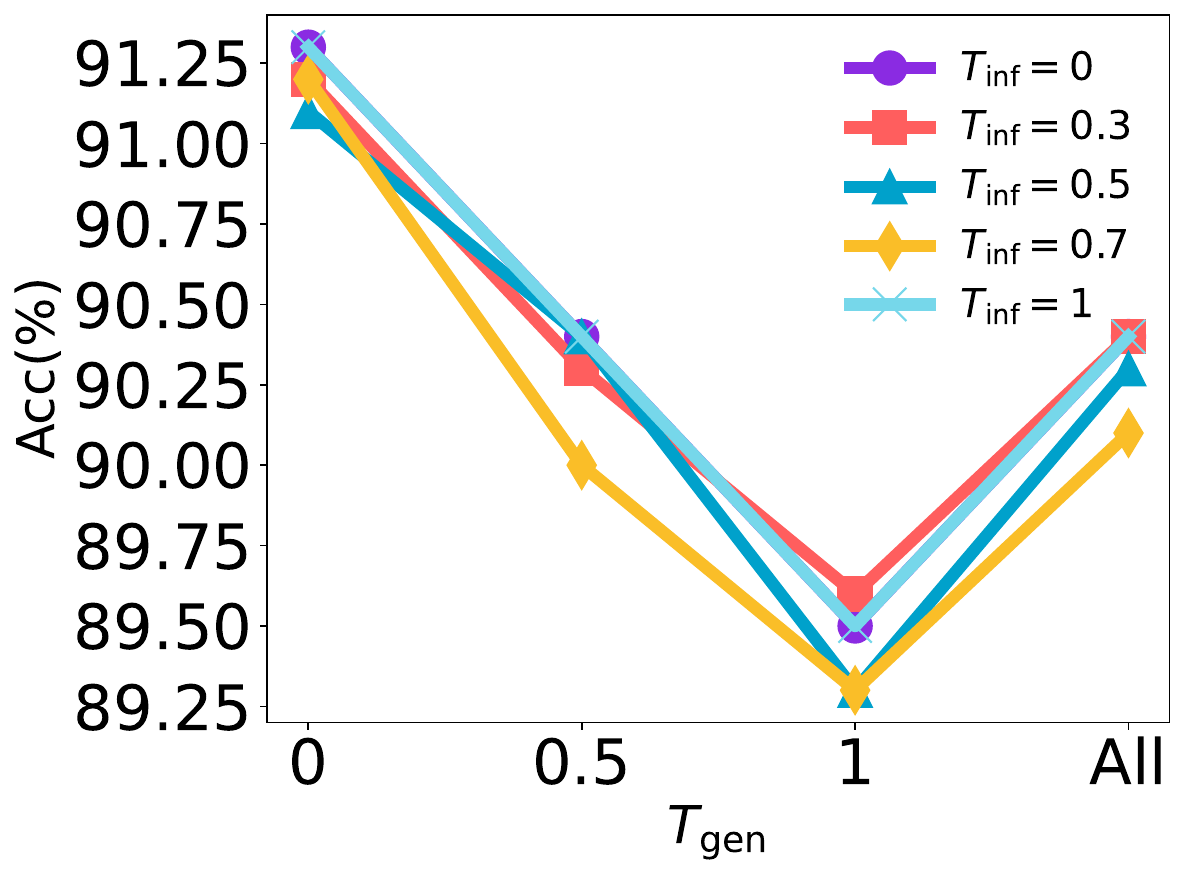}
    \subcaption{}
\end{minipage}\hfill
\begin{minipage}[b]{0.24\textwidth}
    \includegraphics[width=\textwidth]{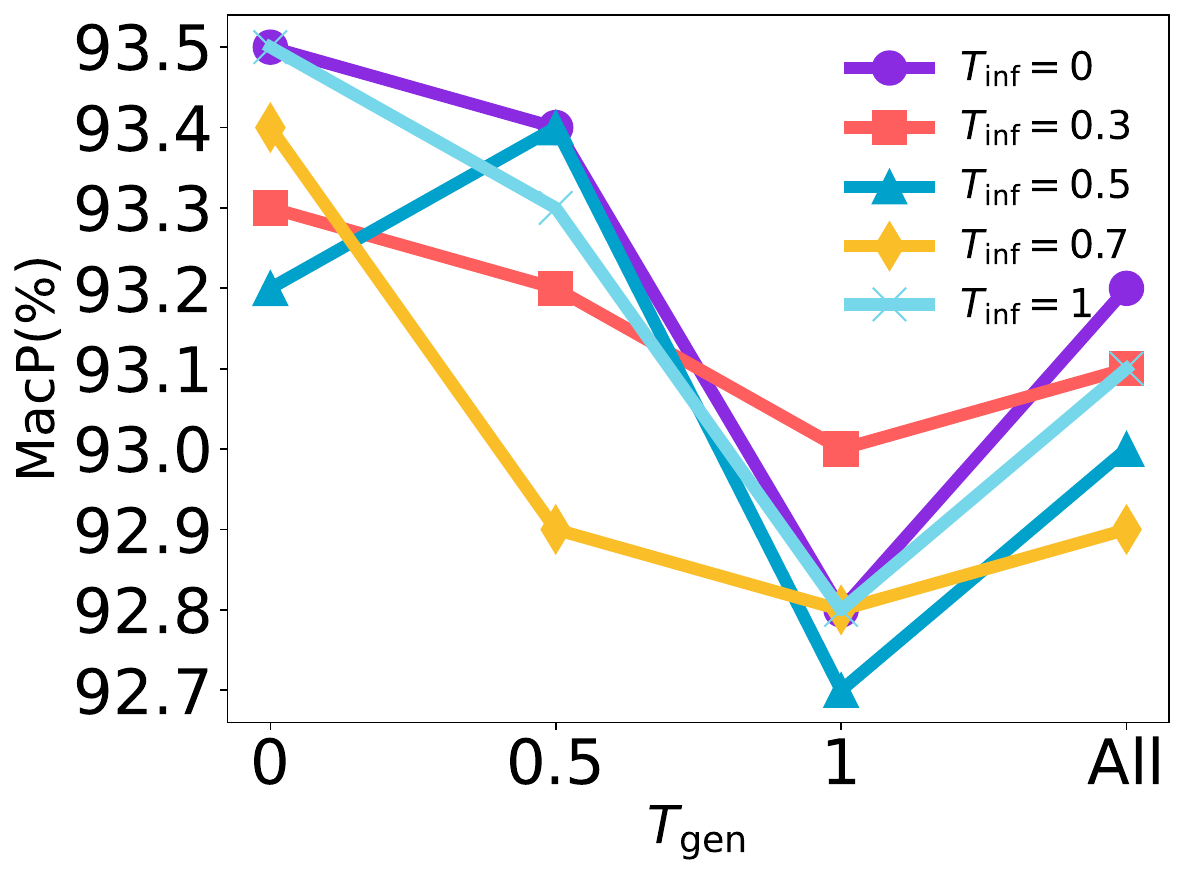}
    \subcaption{}
\end{minipage}\hfill
\begin{minipage}[b]{0.24\textwidth}
    \includegraphics[width=\textwidth]{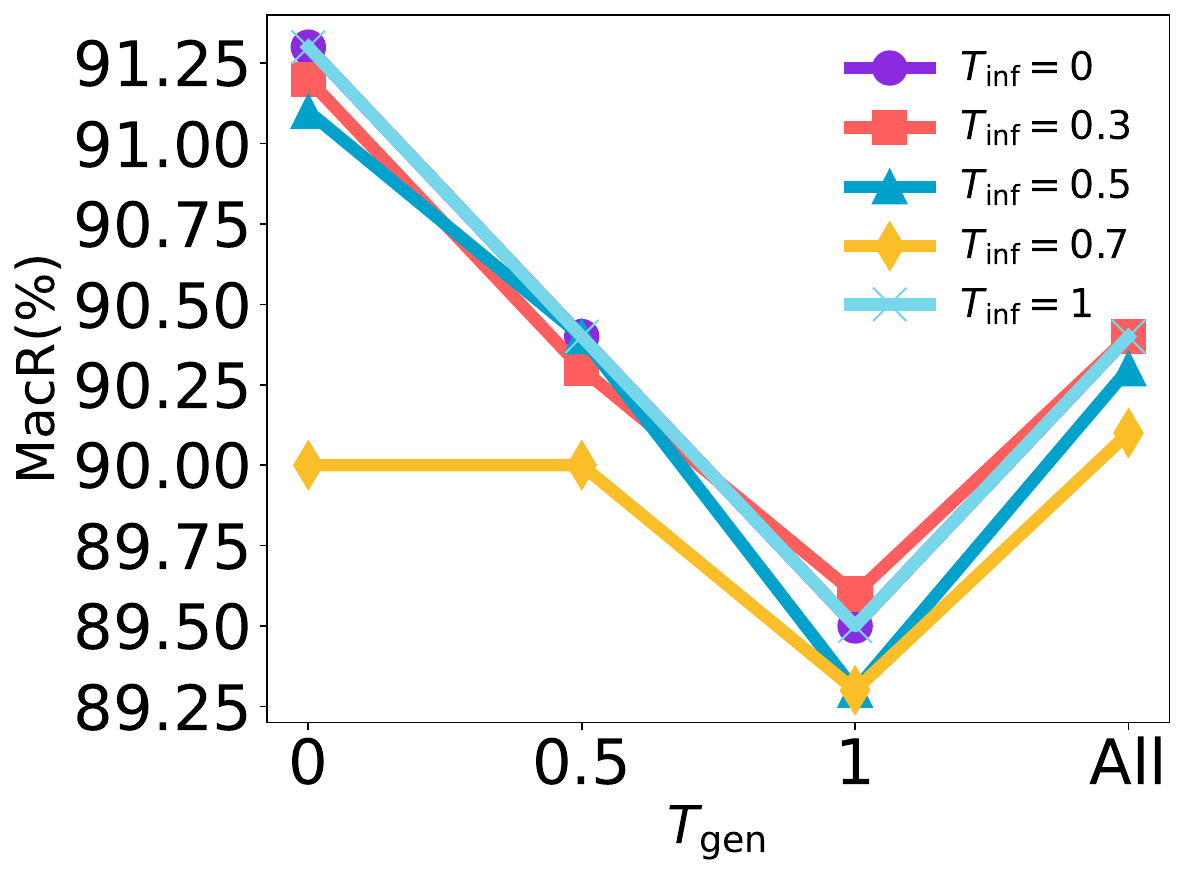}
    \subcaption{}
\end{minipage}\hfill
\begin{minipage}[b]{0.24\textwidth}
    \includegraphics[width=\textwidth]{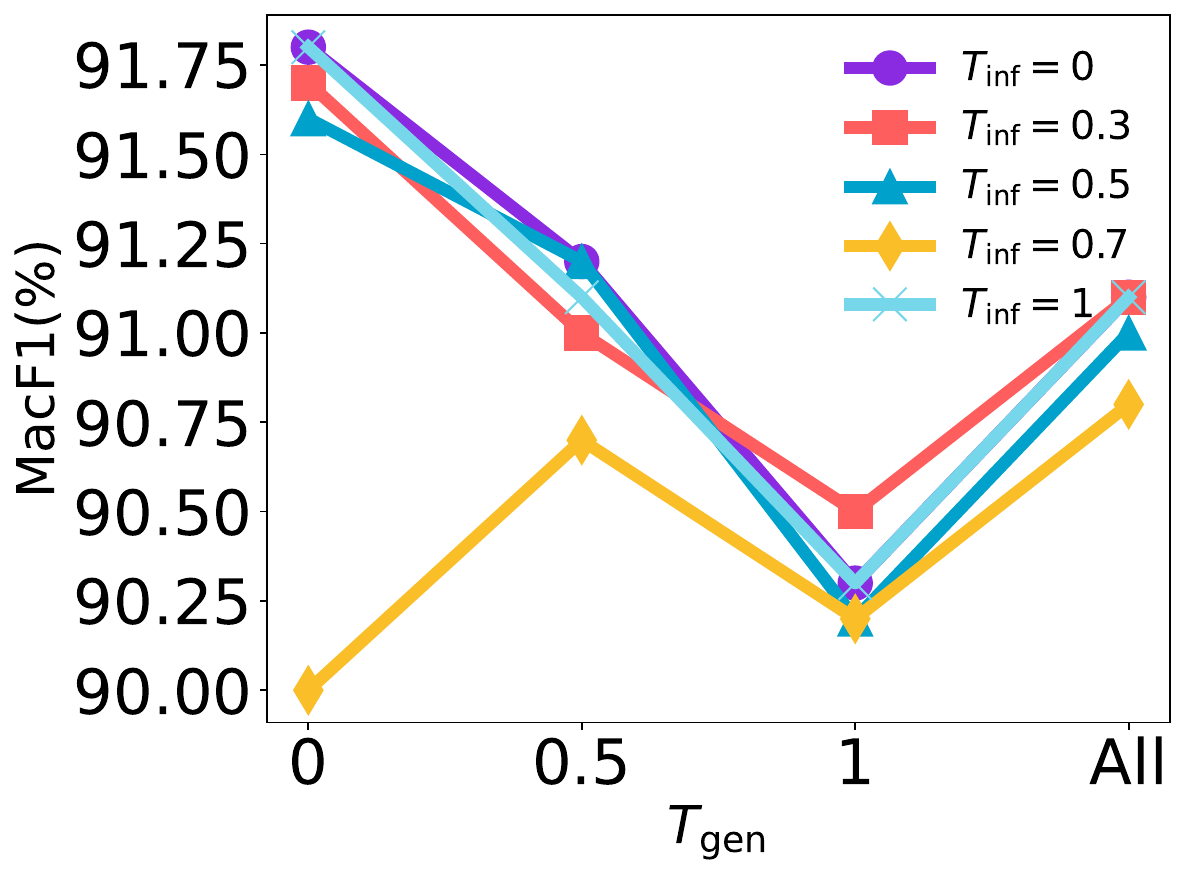}
    \subcaption{}
\end{minipage}
\caption{\NAME accuracy (a), Macro Precision (b), Macro Recall (c), and Macro F1 (d) as functions of generation temperature.  Each curve corresponds to a distinct inference temperature $T_\text{inf}\!\in\!\{0,0.3,0.5,0.7,1\}$; All represents the pooled evaluation set.}
\label{fig:model_metrics_variation}
\end{figure*}
\label{sec:rq5}
To evaluate how hyperparameter choices influence the performance of \NAME, we examine four key factors: (1) training set size, (2) generation and inference temperature, (3) LoRA adapter configuration, and (4) the choice of target modules for LoRA adaptation.

\noindent\textbf{Training Data Size and Generation Temperature.}
We assess model efficiency by training \NAME with 25\%, 50\%, and 100\% of the dataset. As reported in Table~\ref{tab:fdllm_comparison}, \NAME achieves robust performance even with just 25\% of training data: the Macro F1 remains as high as 90.88\% for $T_\text{gen}=0$ and 89.89\% on the full mixed set. Notably, the 50\% setting achieves the highest overall performance, with a Macro F1 of 93.26\% at $T_\text{gen}=0$, even surpassing the 100\% data setting (which achieves 91.88\%), suggesting a slight overfitting effect with the full dataset.

Figure~\ref{fig:model_metrics_variation} further reveals that lower generation temperatures ($T_\text{gen}=0$) lead to higher accuracy and more stable predictions (e.g., 93.20\% at 50\% data, compared to 92.10\% at $T_\text{gen}=1$), while the inference temperature has only a marginal impact. This demonstrates \NAME’s stability across different test-time settings.
\begin{table}[t]
\caption{\NAME Performance Across Different Training Data Proportions.  Each block lists accuracy and macro-metrics for three generation temperatures ($T_\text{gen}\!\in\!\{0,0.5,1\}$) and the joint evaluation set (\textit{All}).}
\label{tab:fdllm_comparison}
\begin{center}
\setlength{\tabcolsep}{5pt}
\begin{tabular}{cccccc}
\toprule
\textbf{Train(\%)} & \textbf{$T_\text{gen}$} & \textbf{Acc(\%)} & \textbf{MacP(\%)} & \textbf{MacR(\%)} & \textbf{MacF1(\%)} \\
\midrule
\multirow{4}{*}{\textbf{25}} & 0     & \textbf{90.67} & \textbf{91.70} & \textbf{90.68} & \textbf{90.88} \\
                             & 0.5   & 89.60          & 91.13          & 89.60          & 89.93          \\
                             & 1     & 88.60          & 90.39          & 88.60          & 88.89          \\
                             & All & 89.62   & 90.90          & 89.62          & 89.89          \\
\hdashline
\multirow{4}{*}{\textbf{50}} & 0     & \textbf{93.20} & \textbf{93.64} & \textbf{93.20} & \textbf{93.26} \\
                             & 0.5   & 92.83          & 93.35          & 92.83          & 92.90          \\
                             & 1     & 92.10          & 92.87          & 92.10          & 92.22          \\
                             & All & 92.71          & 93.20          & 92.71          & 92.79          \\
\hdashline
\multirow{4}{*}{\textbf{100}}& 0     & \textbf{91.33} & \textbf{93.55} & \textbf{91.33} & \textbf{91.88} \\
                             & 0.5   & 90.40          & 93.35          & 90.40          & 91.14          \\
                             & 1     & 89.52         & 92.89         & 89.53          & 90.38          \\
                             & All & 90.42          & 93.19          & 90.42          & 91.12          \\
\bottomrule
\end{tabular}
\end{center}
\end{table}

\begin{table}[t]
  \caption{\NAME performance under different LoRA rank ($r$) and scaling factor ($\alpha$).}
  \label{tab:lora_params}
  \centering
  \begin{tabular}{cccccc}
    \toprule
    \textbf{$r$} & \textbf{$\alpha$} & \textbf{Acc (\%)} & \textbf{MacP (\%)} & \textbf{MacR (\%)} & \textbf{MacF1 (\%)} \\ \midrule
    \multirow{4}{*}{\textbf{256}} & 512 & 88.56 & 86.58 & 84.34 & 84.68 \\
                                  & 256 & 91.48 & 92.76 & 91.48 & 91.60 \\
                                  & \textbf{128} & \textbf{93.27} & \textbf{93.79} & \textbf{93.27} & \textbf{93.28} \\
                                  & 64  & 91.23 & 92.15 & 91.23 & 91.24 \\ \hdashline
    \multirow{2}{*}{128}          & 128 & 92.37 & 88.84 & 87.97 & 88.02 \\
                                  & 64  & 90.99 & 92.18 & 90.99 & 91.05 \\ \hdashline
    \multirow{2}{*}{64}           & 64  & 92.44 & 92.64 & 92.44 & 92.36 \\
                                  & 32  & 89.39 & 86.39 & 85.13 & 85.20 \\ \hdashline
    32                            & 16  & 85.86 & 87.16 & 85.86 & 85.80 \\ \bottomrule
  \end{tabular}
\end{table}

\noindent\textbf{LoRA Parameterization.}
We then examine the effect of the LoRA adapter rank $r$ and the scaling factor $\alpha$. As shown in Table~\ref{tab:lora_params}, the best results are achieved with $r=256$ and $\alpha=128$, giving a Macro F1 of 93.28\%. Configurations where $\alpha = r/2$ (such as $r=256$, $\alpha=128$) generally perform best in our experiments. This trend differs from the common LoRA recommendation of setting $\alpha = 2r$. We also observe that a large $\alpha$ (for example, $r=256$, $\alpha=512$, Macro F1 = 84.68\%) leads to reduced performance. Low rank values limit adaptability (e.g., $r=32$, $\alpha=16$, Macro F1 = 85.80\%).
Notably, following the standard setting does not yield optimal results in our task. 
Our findings suggest that the optimal parameterization for attribution is not the same as for conventional downstream tasks. This difference may reflect the unique characteristics of LLM fingerprinting, where capturing implicit generation patterns requires different capacity and regularization trade-offs.

\begin{table}[t]
  \caption{Performance difference relative to LoRA baseline.}
  \label{tab:peft_diff}
  \centering
  \setlength{\tabcolsep}{2pt}
  \begin{tabular}{ccccc}
    \toprule
    \textbf{Metric} & \textbf{AdaLoRA~\cite{zhang2023adaptive}} & \textbf{QLoRA~\cite{dettmers2023qlora}} & \textbf{DoRA~\cite{dora2024}} & \textbf{LoRA$+$~\cite{hayou2024lora+}} \\ \midrule
    \textbf{ACC(\%)}   & -9.35 & -2.96 & -2.59 & -0.15 \\
    \textbf{MacF1(\%)} & -9.38 & -2.95 & -2.56 & -0.15 \\ \bottomrule
  \end{tabular}
\end{table}

\noindent\textbf{Comparison with Other PEFT Strategies.}
As summarized in Table~\ref{tab:peft_diff}, we compare its improved variant LoRA$+$ and several recent PEFT methods~\cite{peftguard}, including AdaLoRA and QLoRA, under consistent experimental conditions. LoRA and LoRA$+$ outperform AdaLoRA and QLoRA in all main metrics. For instance, AdaLoRA shows a decrease of 9.38\% in Macro F1 compared to LoRA, while LoRA$+$ achieves nearly identical results to LoRA (less than 0.2\%). 
This suggests that vanilla LoRA remains both reliable and efficient for our attribution setting. 

\begin{table}[t]
\centering
\caption{Performance of \NAME on Different Target Modules. \textbf{P} indicates the proportion of trainable parameters.}
\label{table:target_modules}
\begin{tabular}{cccc}
\toprule
\textbf{Target Modules} & \textbf{P(\%)} & \textbf{Acc(\%)} & \textbf{MacF1(\%)} \\
\midrule
$[\Delta w_q]$ & 0.0464 & 87.42  & 87.41 \\
$[\Delta w_q, \Delta w_k]$ & 0.0723 & 87.70 & 87.69 \\
$[\Delta w_q, \Delta w_k, \Delta w_v]$ & 0.0982 & 89.00 & 89.03 \\
$[\Delta w_q, \Delta w_k, \Delta w_v, \Delta w_o]$ & 0.1435 & 90.25 & 90.25 \\ \hdashline
$[\Delta w_{\text{gate}}]$ & 0.1435 & 92.23 & 92.22 \\
$[\Delta w_{\text{up}}]$ & 0.1435 & 91.32 & 91.12 \\
$[\Delta w_{\text{down}}]$ & 0.1435 & 89.56 & 89.55 \\
$[\Delta w_{\text{up}},\Delta w_{\text{down}}]$ & 0.2857 & 91.51 & 91.50 \\
$[\Delta w_{\text{gate}}, \Delta w_{\text{up}}, \Delta w_{\text{down}}]$ & 0.4272 & 92.43 & 92.43 \\
\bottomrule
\end{tabular}
\end{table}

\noindent\textbf{Comparison of Performance across Target Modules.}
Table~\ref{table:target_modules} compares the outcomes of selectively fine-tuning different model components. When only the attention projections are tuned, the best accuracy (90.25\%) and Macro F1 (90.25\%) are achieved by updating all four attention matrices. Notably, even when only the $\Delta w_\text{gate}$ in the feedforward network (FFN) is tuned, the model achieves strong results, with 92.23\% accuracy and 92.22\% Macro F1. Jointly adapting all three FFN layers further improves performance, reaching 92.43\% for both accuracy and Macro F1.

This result differs from common LoRA applications, where fine-tuning attention modules typically yields the best gains on conventional downstream tasks. In our attribution setting, adapting the FFN modules yields significantly higher accuracy. This finding suggests that attribution tasks rely more on implicit changes in the LLM’s internal representations, which are more effectively captured by the feedforward layers. 

\begin{tcolorbox}
\textbf{Take-aways:}
Our experiments demonstrate that \NAME achieves robust and stable attribution performance. Collecting texts generated under different temperature settings further enhances detection accuracy. Lower inference temperatures generally yield higher attribution scores. LoRA’s best results are obtained with moderate rank and scaling factors, while overparameterization can be detrimental. Fine-tuning FFN layers provides greater improvements than attention-only tuning, enabling effective adaptation with minimal trainable parameters. 
\end{tcolorbox}

\section{Limitations \& Future Work}\label{sec:discussion}

Our study relies on data collected between September 2024 and May 2025, which imposes temporal limitations. Future work will focus on improving the initial training data construction to enhance \NAME's generalization and robustness in real-world scenarios.
Given the rapid evolution of LLMs, it was not feasible to include all mainstream models in our evaluation. We did not test our approach on newly released models such as Gemini2.5~\cite{gemini2.5} and Claude 4~\cite{claude4}.
To maintain the practical relevance of \NAME for real-world LLMGT detection, we plan to conduct regular model updates and will continue to make our research publicly available.

We currently focus on text generated by single LLMs or LLM pairs. However, as multi-agent systems mature, user workflows increasingly involve multiple collaborating LLMs. This creates more complex detection scenarios where outputs reflect multiple distinct models. Since LLMs remain foundational to most agent-based systems, our method provides a solid basis for future analysis in such settings.

\section{Related Work}\label{sec:relatedwork}

Previous studies have primarily focused on the task of Authorship Attribution (AA) \cite{uchendu2020authorship,abdali2024decoding}. Uchendu et al.\cite{uchendu2023attribution} highlighted the limitations of traditional AA methods and proposed shifting the focus to Authorship Attribution for Neural Text Generators. 
They classified existing AA approaches as stylometric, deep learning, statistical, and hybrid, concluding that deep learning methods perform best for AA tasks. Recent research on LLM attribution generally follows two directions: white-box approaches that modify model parameters or outputs and black-box approaches that rely solely on generated text.

\textbf{White-Box Techniques}. In white-box settings, LLM identification often relies on watermarking~\cite{cohen2024watermarking}, which embeds identity via algorithmic word substitutions~\cite{bahri2024watermark}. Xu et al.\cite{xu2024freqmark} proposed periodic signal-based watermarks, while Google’s SynthID-Text\cite{dathathri2024scalable} modifies the sampling process for scalable detection. However, these methods require altering model parameters or sampling strategies, which may degrade performance or affect output quality.

\textbf{Black-Box Techniques}.  
Without internal access, detection becomes a text-only classification problem, typically divided into two categories: 
\noindent\textit{(1) Metric-Based Methods}.
Early academic research primarily relied on mathematical metrics to distinguish the generated text. These studies typically focused on binary classification, aiming to distinguish between human-written content and machine-generated text. Pre-LLM approaches often leveraged straightforward cues such as token probabilities, rank histograms, or entropy. For example, GLTR~\cite{gehrmann2019gltr} visualizes the probability mass left by the generator, while Solaiman et al.~\cite{solaiman2019releasestrategiessocialimpacts} use the average word-level log-likelihood score to judge a text.
\noindent\textit{(2) Model-Based Methods}.  
Researchers have gradually turned their attention to leveraging the models, utilizing their powerful learning abilities to determine model identities~\cite{yang2023dna}. For instance, Zeng et al.~\cite{zeng2023huref} employed CNNs to learn invariants in model parameters and used StyleGAN2 to generate human-recognizable natural images. Similarly, Li et al.~\cite{li2024mage} evaluated four main detection methods, including supervised approaches (e.g., classifiers based on pre-trained language models~\cite{beltagy2020longformer}) and unsupervised approaches (e.g., DetectGPT~\cite{mitchell2023detectgpt}) to distinguish between human- and machine-generated text. Their findings demonstrated that detection methods based on deep learning models perform well in binary classification tasks. Shi et al.~\cite{Shi_2024} proposed a method for detecting LLMs by repeatedly resampling to extract text features, simulating white-box detection in a black-box environment.

\section{Conclusion}\label{sec:conclusion}

We first introduce \textit{FD-Dataset}, a bilingual dataset comprising 90,000 samples generated by 20 advanced LLMs. It is specifically designed to support robust fingerprinting evaluation under black-box conditions.
Building upon this dataset, we propose \NAME, a novel detector that employs LoRA-based adaptation to capture subtle features. 
Extensive experiments show that \NAME consistently achieves high attribution accuracy across various scenarios, including unseen models and adversarial attacks.
Through an analysis of the attribution mechanism, we reveal why LoRA is particularly effective in this task: LoRA adaptation enables the foundation model to capture deep, persistent features that aggregate the same LLM while enhancing the separation between different LLMs. 
\NAME maintains strong performance against unseen models and OOD samples.
\NAME is also much more robust than prior methods against attacks such as polishing, translation, and synonym substitution.
We hope our research will support the responsible use of LLMGT.

\newpage



\bibliographystyle{IEEEtran}
\bibliography{bare_conf_compsoc}

\appendices

\section{Data Construction Prompts}
\label{appendixprompts}
\begin{figure}[htbp]
    \centering
    \begin{subfigure}[b]{0.48\columnwidth}
        \centering
        \includegraphics[width=\textwidth]{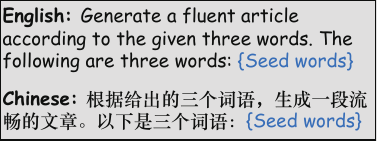}
        \caption{Generation Prompt}
        \label{fig3a}
    \end{subfigure}
    \begin{subfigure}[b]{0.48\columnwidth}
        \centering
        \includegraphics[width=\textwidth]{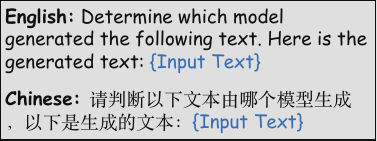}
        \caption{Inference Prompt}
        \label{fig3b}
    \end{subfigure}
    \caption{Prompt design examples.}
    \label{fig:combined}
\end{figure}
Figure~\ref{fig:combined} presents two representative examples of the prompt templates used in our data construction pipeline:

\begin{itemize}
    \item \textbf{Generation Prompt}: This template is employed during the dataset creation phase. It is designed to encourage LLMs to produce natural, diverse, and high-coverage text samples across various topics and domains. The prompt formulation minimizes bias and helps capture distinct generation patterns unique to each model.
    \item \textbf{Inference Prompt}: This template is used at the training and evaluation stage. It serves to test the model's response characteristics under consistency.
\end{itemize}

\section{Alternative Loss Formulation}
\label{appendix:aco}
As an alternative, we also considered a contrastive loss formulation that explicitly structures the latent space by minimizing intra-class distances (among samples from the same source LLM) and maximizing inter-class distances (among samples from different LLMs). Specifically, for an anchor sample $i$, a positive sample $j$ from the same class, and a negative sample $k$ from a different class, the contrastive objective is abstractly written as:
\begin{align}
\min_{\Delta\mathbf{W}} \Bigg\{
    \sum_{i} 
    & \underbrace{\sum_{j \in C_{y_i}} \left\| \mathbf{F}'_i - \mathbf{F}'_j \right\|^2}_{\text{intra-class distance}} \notag \\
    & - \lambda \underbrace{\sum_{i} \sum_{k \notin C_{y_i}} \left\| \mathbf{F}'_i - \mathbf{F}'_k \right\|^2}_{\text{inter-class distance}}
\Bigg\}
\label{eq:contrastive_lora}
\end{align}
where $\lambda$ balances the two terms. While this loss can structure the representation space more explicitly, we found the CE loss to be more effective and practical for our setting and, therefore, adopted it for all reported experiments.

\section{Overview of Centroid}
\label{appedix:oc}
\begin{figure}[t]
\centering 
\includegraphics[width=\linewidth]{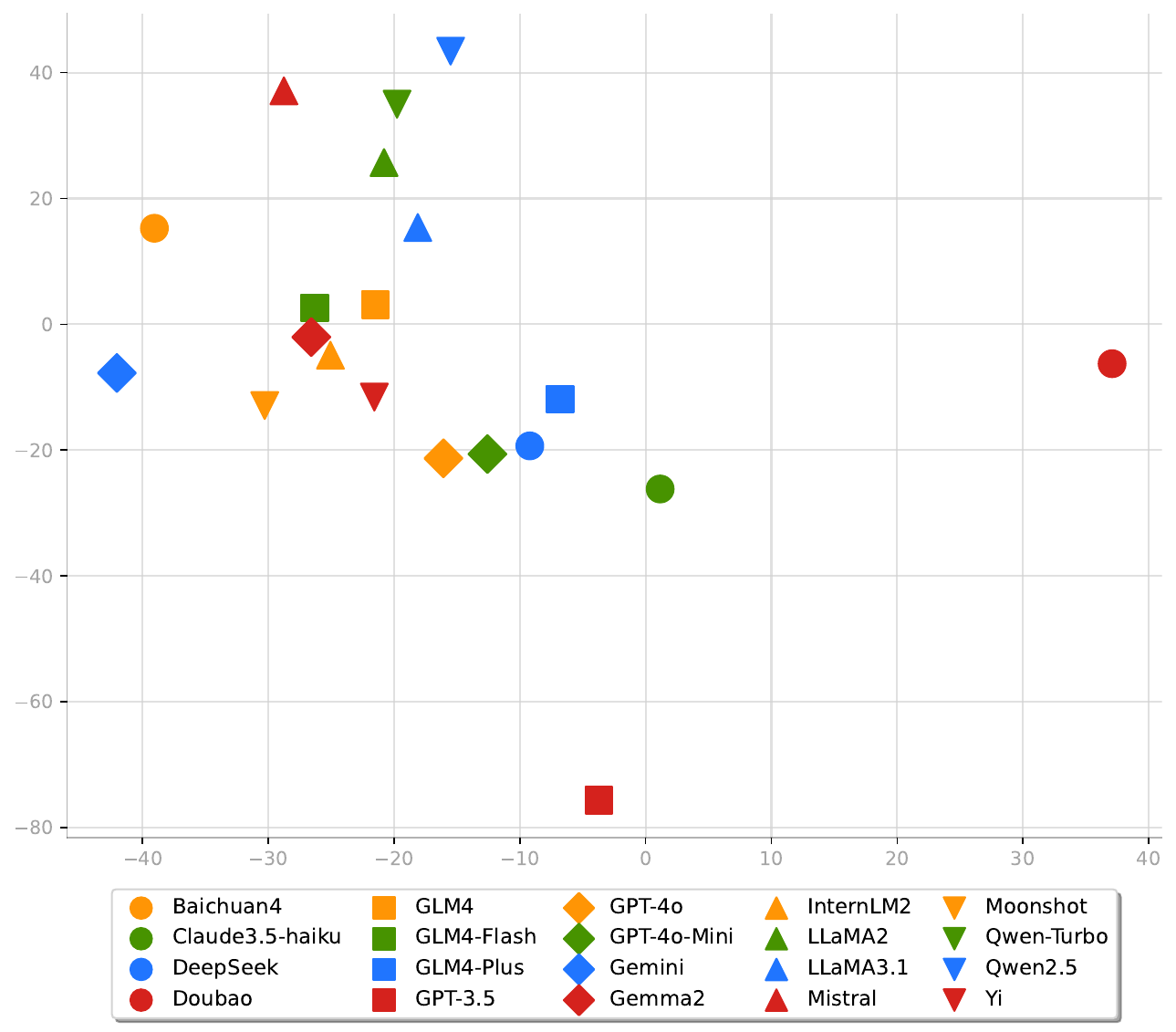}
\caption{Two-dimensional projection of 20 advanced LLMs in the FDLLM fingerprint space.}
\label{fig:oc}
\end{figure}

In Figure~\ref{fig:oc} each dot represents a model, and its color and marker encode the model. Two key patterns emerge:
(1) Points for the same family are tightly grouped, indicating that FDLLM captures highly consistent features.
(2) Models such as GPT-3.5 and Yi lie far from the main cluster, showing that FDLLM can distinguish differences in architecture or training strategies.

\section{LLMs}
\label{llms}
\textbf{GPT-3.5~\cite{GPT-3.5-Turbo}.} GPT-3.5 represents a family of LLMs developed and released by OpenAI. As part of the Generative Pre-trained Transformer (GPT) series, GPT-3.5 builds upon the architecture and capabilities of the original GPT-3 models.

\textbf{GPT-4o~\cite{openai2024gpt4technicalreport}.} GPT-4o is OpenAI's flagship multimodal model, representing a significant step towards more natural human-computer interaction. The "o" stands for "omni," highlighting its native ability to seamlessly process and generate content across text, audio, and visual modalities within a unified model. 

\textbf{GPT-4o-mini~\cite{openai2024gpt4technicalreport}.} GPT-4o-mini is a smaller, highly efficient variant derived from the GPT-4o architecture, developed by OpenAI. It is designed to offer a compelling balance between performance, speed, and affordability. 

\textbf{Gemini-1.5~\cite{team2024gemini}.} Gemini-1.5 is a competent multimodal model developed by Google DeepMind. It utilizes a Mixture-of-Experts (MoE) architecture, contributing to enhanced efficiency and performance across a wide range of complex reasoning tasks involving text, code, audio, image, and video modalities.

\textbf{Gemma2~\cite{gemmateam2024gemma2improvingopen}.} Gemma2 represents the next generation of lightweight, building upon the technology used to create the Gemini models. Available in various sizes (e.g., 9B and 27B parameters), Gemma2 offers researchers and developers powerful yet accessible open-weight models suitable for a wide range of applications.

\textbf{Claude3.5~\cite{claude35haiku}.} Claude 3.5 demonstrates marked improvements in understanding nuance, humor, and complex coding problems while excelling at generating high-quality, natural-sounding content. It also features substantial advancements in visual reasoning capabilities compared to previous Claude models. It operates with significantly enhanced speed and cost-effectiveness, aiming to deliver top-tier intelligence more broadly.

\textbf{Llama2~\cite{touvron2023llama2openfoundation}.} Llama 2 was designed as an open-source resource available for both research and commercial use. The family includes base models and fine-tuned chat versions optimized for dialogue use cases through supervised fine-tuning and reinforcement learning with human feedback (RLHF). 

\textbf{Llama3.1~\cite{grattafiori2024llama3herdmodels}.} Llama3.1 represents the next iteration of Meta's open-source LLM series. This release introduces several models, notably powerful 8B, 70B, and a new state-of-the-art 405B parameter version, trained on significantly larger and more diverse datasets.

\textbf{Qwen-turbo~\cite{qwen25}.} Qwen-turbo is a specific LLM variant within the Qwen LLM family developed by Alibaba Cloud. It is typically optimized for speed and efficiency and designed to respond rapidly to applications requiring low latency.

\textbf{Qwen2.5~\cite{qwen25}.} Qwen2.5 represents a significant update to the Qwen series of LLMs from Alibaba Cloud. This generation introduces improvements across various capabilities, including enhanced language understanding, reasoning, coding, and multimodal processing. 

\textbf{Qwen3~\cite{qwen3}.} Qwen3 is the newest generation in the Qwen family of large-language-model suites, offered in both dense and MoE variants. A single model can fluidly switch between "thinking mode" (optimized for demanding reasoning, math, and coding) and "non-thinking mode" (fast, general-purpose dialogue), delivering top-tier results across tasks. 

\textbf{GLM4~\cite{glm2024chatglm}.} GLM4 is the fourth generation of the General Language Model (GLM) series, developed collaboratively by Zhipu AI and Tsinghua KEG. GLM-4-9B is an open-source version of the GLM-4 series, which excels in semantics, mathematics, reasoning, code, and knowledge. GLM-4-Flash and GLM-4-Plus are proprietary closed-source versions of the GLM-4 series.

\textbf{Deepseek~\cite{deepseekai2024deepseekv3technicalreport}.} DeepSeek is typically designed with innovative architectures (like Mixture-of-Experts in V2) to achieve strong performance while aiming for training efficiency. Trained on 8.1 trillion diverse, high-quality tokens, DeepSeek-V2 undergoes Supervised fine-tuning and Reinforcement Learning stages to thoroughly realize its capabilities.

\textbf{InternLM2~\cite{cai2024internlm2}.} InternLM2 is an open-source LLM that surpasses its predecessors through innovative pre-training and optimization techniques. The model demonstrates comprehensive performance enhancements across various capabilities, including reasoning, mathematics, and coding.

\textbf{Moonshot~\cite{mootshot}.} Moonshot is a language model with hundreds of billions of parameters launched by Moonshot AI. The Moonshot model can be applied to various tasks, including content and code generation, summarization, and creative writing.

\textbf{Doubao~\cite{doubao}.} Doubao represents ByteDance's significant investment in generative AI. It is understood to power various applications within the ByteDance ecosystem. It is likely optimized for performance across diverse tasks, potentially with a strong focus on the Chinese language and multimodal capabilities.

\textbf{Yi~\cite{ai2024yi}.} Yi series models are the next generation of open-source LLMs trained from scratch by 01.AI. Targeted as a bilingual language model and trained on a 3T multilingual corpus, the Yi series models become one of the strongest LLMs worldwide, showing promise in language understanding, commonsense reasoning, reading comprehension, and more.

\textbf{Mistral~\cite{jiang2023mistral}.} Mistral marked a significant step in efficient LLM design. Despite its relatively small size (7B parameters), Mistral demonstrated remarkable performance, outperforming larger models on numerous reasoning, mathematics, and code generation benchmarks.

\textbf{Baichuan4~\cite{baichuan4}.} Baichuan4 represents the latest generation of LLMs from Baichuan Intelligence Technology. As part of the Baichuan series, which has often included open-source contributions, Baichuan4 likely represents a state-of-the-art model within the Chinese AI landscape, reflecting rapid advancements in the field.

\textbf{Phi4~\cite{phi4}.} Phi-4 is a 14B parameters language model whose standout strengths come from a data-centric training recipe: high-quality synthetic data is woven into pre-training, curriculum design, and post-training. With only minor architectural tweaks beyond Phi-3, these improvements enable Phi-4 to outperform its teacher (GPT-4) in STEM-oriented reasoning and achieve top-tier results for its size.

\textbf{Granite3.3~\cite{granite33}.} Granite 3.3 is an open-weight LLM tuned with permissively licensed instructions and long-context synthetic data. It upgrades earlier versions with stronger reasoning and Fill-in-the-Middle code completion while retaining 128 K context, robust RAG/function-calling, and tunable length/creativity controls, achieving competitive scores on general, enterprise, and safety benchmarks.

\textbf{MiMo~\cite{xiaomi2025mimo}.} MiMo is a scratch-trained, reasoning-centric 7B parameters model whose reinforcement-learning fine-tuned variant surpasses typical 32 B models and rivals OpenAI o1-mini on both math and code reasoning tasks.

\end{document}